\documentclass[prd,preprint,superscriptaddress,preprintnumbers,eqsecnum,showpacs,nofootinbib,nobibnotes,noeprint,floatfix,dvipsnames]{revtex4-1}
\usepackage{amsfonts,amsmath,mathbbol,amssymb,bm,natbib}
\usepackage[table,dvipsnames]{xcolor}
\usepackage{graphicx} 
\usepackage{color}
\usepackage{stackengine}
\usepackage{scalerel}
\usepackage{mathtools}
\usepackage[linkcolor=blue,citecolor=blue,urlcolor=blue,colorlinks=true]{hyperref}
\definecolor{refkey}{rgb}{0.39,0.58,1}
\definecolor{labeled}{rgb}{1,0,0}

\usepackage{tabularx}
\usepackage{multirow}
\usepackage{mathrsfs}
\usepackage{array}

\usepackage{float}
\usepackage{amsmath}
\usepackage{physics}

\def\ie{{\it i.e.}, }
\def\eg{{\it e.g.}, }
\newcommand{\be}{\begin{equation}}
\newcommand{\bea}{\begin{eqnarray}}
\newcommand{\ee}{\end{equation}}
\newcommand{\eea}{\end{eqnarray}}

\def\s#1{{\scriptscriptstyle #1}}
\def\srm#1{{\rm{\scriptscriptstyle #1}}}

\def\1eq#1{Eq.~(\ref{#1})}

\def\2eqs#1#2{Eqs.~(\ref{#1}) and~(\ref{#2})}
\def\3eqs#1#2#3{Eqs.~(\ref{#1}),~(\ref{#2}) and~(\ref{#3})}
\def\6eqs#1#2#3#4#5#6{Eqs.~(\ref{#1}),~(\ref{#2}),~(\ref{#3}),~(\ref{#4}),~(\ref{#5}) and~(\ref{#6})}

\def\fig#1{Fig.~\ref{#1}}

\newcommand{\gp}{{\bm{\Gamma}}}
\newcommand{\gh}{\widehat{\bm{\Gamma}}}
\newcommand{\gt}{\widetilde{\bm{\Gamma}}}
\newcommand{\acc}{${\rm Q\bar{c}c}\,$}
\newcommand{\bbcc}{${\rm BB\bar{c}c}\,$}
\newcommand{\bqcc}{${\rm BQ\bar{c}c}\,$}
\newcommand{\bcc}{${\rm B\Bar{c}c}\,$}
\newcommand{\bqq}{${\rm BQQ}\,$}
\newcommand{\bbqq}{${\rm BBQQ}\,$}

\newcommand{\cc}{\overline{c}}

\newcommand{\zt}{\widetilde{Z}}

\def\srm#1{{\rm{\scriptscriptstyle #1}}}

\begin{document}

\title{Schwinger-Dyson truncations in the all-soft limit: a case study}

\author{A.~C.~Aguilar}
\affiliation{\mbox{University of Campinas - UNICAMP, Institute of Physics ``Gleb Wataghin'',} 
13083-859 Campinas, S\~{a}o Paulo, Brazil.}

\author{M.~N. Ferreira}
\affiliation{\mbox{Department of Theoretical Physics and IFIC,} \\ University of Valencia and CSIC, E-46100, Valencia, Spain.}

\author{B.~M. Oliveira}
\affiliation{\mbox{University of Campinas - UNICAMP, Institute of Physics ``Gleb Wataghin'',} 13083-859 Campinas, S\~{a}o Paulo, Brazil.}

\author{J.~Papavassiliou}
\affiliation{\mbox{Department of Theoretical Physics and IFIC,} \\ University of Valencia and CSIC, E-46100, Valencia, Spain.}


\begin{abstract}

We study a special Schwinger-Dyson equation in the context of a 
pure SU(3) Yang-Mills theory, formulated in the 
background field method.
Specifically, we consider the corresponding equation for 
the vertex that governs the interaction of two background 
gluons with a ghost-antighost pair.   
By virtue of the background gauge invariance, this vertex 
satisfies a naive Slavnov-Taylor identity, 
which is not deformed by the ghost sector of the theory.  
In the all-soft limit, where all momenta vanish,  
the form of this vertex may be obtained exactly from
the corresponding Ward identity.
This special result is subsequently reproduced 
at the level of the Schwinger-Dyson equation,
by making extensive use of Taylor's theorem and exploiting a plethora of key 
relations, particular to the background field method.
This information permits the determination of the error associated 
with two distinct truncation schemes, where the potential advantage from employing
lattice data for the ghost dressing function  
is quantitatively assessed. 
\end{abstract}

\maketitle

\section{Introduction}
\label{sec:intro}

The Schwinger-Dyson equations (SDEs) form 
an infinite tower of coupled non-linear integral equations
that govern the dynamical evolution of 
all $n$-point Green's (correlation) functions of a quantum field theory~\cite{Dyson:1949ha,Schwinger:1951ex,Schwinger:1951hq}. 
The SDEs are derived formally from the generating functional of the theory~\cite{Rivers:1987hi,Itzykson:1980rh}, and constitute one of the few nonperturbative frameworks 
available in the continuum~\cite{Roberts:1994dr,Alkofer:2000wg,Maris:2003vk,Fischer:2006ub, Binosi:2009qm,Binosi:2008qk,Huber:2018ned}. Over the years they 
have been employed in the study of a wide array of physical phenomena, encompassing, among others,  
superconductivity~\cite{Dorey:1990sz, Dorey:1991kp,Lee:2002qza,Popovici:2013fya}, dynamical chiral symmetry breaking~\cite{Aguilar:2018epe,Gao:2021wun,Mitter:2014wpa,Aguilar:2010cn,Fischer:2003rp,Roberts:1994dr}, and the emergence of mass in strongly coupled theories, such as pure Yang-Mills theories and Quantum Chromodynamics (QCD)~\cite{Cornwall:1981zr,Aguilar:2008xm,Aguilar:2019kxz,Aguilar:2013hoa,Aguilar:2020uqw,Aguilar:2021uwa,Horak:2022aqx,Papavassiliou:2022wrb}.

Even  though, in  principle, the  SDEs encode  the complete  dynamical information of  all correlation functions  of the theory, in practice    their   treatment    requires   the    implementation   of truncations.  For instance,  certain vertices or multi-particle kernels that  
enter in the diagrammatic representation of a given SDE
may be set to  their tree-level  value, or be completely neglected.  Similarly,  dressed-loop approximations may be
adopted,  where only  a given  order  of diagrams  in a  loop-wise
expansion is retained.  However, due the lack of a definite expansion parameter, there is no a-priori way of 
estimating the error committed due to such approximations.
Instead, the errors may  be estimated only  a-posteriori,  either  by direct  comparison  with  experimental results or  lattice simulations, or, more  laboriously, by introducing further structures, \ie dressing vertices or adding loops, and computing their numerical impact.
This is to be contrasted with approaches possessing an obvious expansion parameter, 
such as large $N_c$~\cite{tHooft:1973alw,Witten:1979kh,Coleman1985}, or heavy mass ($M$) expansions~\cite{Eichten:1989zv,Georgi:1990um}, where, at the $n$-th step, the neglected terms 
are of order ${\cal O} (1/N_c^{n+1})$ or  ${\cal O} (1/M^{n+1})$.

It would be clearly instructive to consider a toy SDE scenario
where the exact result for the Green's function in question 
is known by virtue of 
general field-theoretic principles,  
and the numerical impact of certain typical truncations
may be easily evaluated. To that end, 
we turn to the well-known framework of the Background Field Method (BFM)~\cite{DeWitt:1967ub,Honerkamp:1972fd,Kallosh:1974yh,Kluberg-Stern:1974nmx,Arefeva:1974jv,Abbott:1980hw,Weinberg:1980wa,Abbott:1981ke,Shore:1981mj,Abbott:1983zw}, 
where the gauge field $A_\mu$ is decomposed 
as $A_\mu = B_\mu + Q_\mu $, with $B_\mu$ the classical (background) 
part and $Q_\mu$ the quantum (fluctuating) component, 
and a special gauge-fixing procedure is employed. 
Within this formalism we derive and 
analyze the SDE of the four-particle vertex 
that consists of two background gluons and a ghost-antighost pair, to be denoted by \bbcc. It is important to emphasize 
that, due to the 
background gauge symmetry, this vertex satisfies an Abelian 
Slavnov-Taylor identity (STI) when contracted by the momentum 
carried by any of its  background legs. Note that Abelian STIs  
are direct generalizations of tree-level relations, and, in contradistinction to the STIs~\cite{Taylor:1971ff,Slavnov:1972fg} 
of the linear covariant gauges~\cite{Fujikawa:1972fe}, they 
receive no modifications from the ghost sector of the theory. 

It turns out that it is possible to obtain an exact nontrivial result for
the vertex \bbcc by appealing directly to 
this latter STI, which relates the divergence of \bbcc  
with the three-particle vertex $\rm B\cc c$~\cite{Aguilar:2017dco} at different permutations of its arguments.  
As some of the momenta involved are set to zero, certain known limits of the
vertex $\rm B\cc c$ are triggered; and finally, 
in the all-soft limit, \ie when all incoming momenta vanish,
the STI becomes a Ward identity (WI) 
that expresses \bbcc in terms of the ghost-dressing function at the origin. Past its formal simplicity, the main advantage of this results is that, in the Landau gauge, it
fully determines the deep infrared structure of the vertex \bbcc
in terms of a quantity that has been extensively studied
both on the lattice~\cite{Sternbeck:2005tk,Ilgenfritz:2006he,Cucchieri:2007md,Bogolubsky:2007ud,Cucchieri:2008fc,Cucchieri:2009zt,Bogolubsky:2009dc,Maas:2011se,Boucaud:2011ug,Ayala:2012pb,Boucaud:2017ksi,Boucaud:2018xup} and in the continuum~\cite{Aguilar:2008xm, Boucaud:2008ky,Fischer:2008uz,
Tissier:2010ts,Pennington:2011xs,Vandersickel:2012tz,Dudal:2012zx,Aguilar:2013xqa,Cyrol:2017ewj,Gao:2017uox,Aguilar:2018csq,Corell:2018yil,Huber:2018ned,Aguilar:2021okw}.

Evidently, when all incoming momenta are 
set to zero at the level of the SDE governing the \bbcc, and 
in the absence of truncations or approximations, 
\ie when the SDE is treated exactly, 
the above result must emerge identically. 
However, as we will elucidate in the main text, the 
correct implementation of the all-soft limit is rather subtle, 
hinging on fundamental properties 
of vertices and kernels entering in the diagrammatic expansion 
of the SDE under consideration. 
Once all field-theoretic principles have been correctly taken into account, 
one recovers precisely the same result 
at the level of the SDE as that obtained from the WI. 

The above analysis is particularly instructive, because it exposes the delicate 
interplay required among various components in order to  
preserve fundamental symmetries at the level of SDEs. 
In that sense, the derivation of an exact WI from a vertex SDE,
presented in this article,  
constitutes a rather noteworthy result.
Moreover, the errors induced by certain 
truncations or approximations 
may be estimated by comparing directly the approximate answer 
with the exact result.  
This possibility is 
particularly welcome 
in a SDE context, 
where the absence of a concrete expansion parameter 
obscures the task of  
assigning errors to the results obtained.

In order to explore this last point in detail, 
we consider a concrete truncation, which is rather natural 
in this context, namely we approximate the 
full ghost-gluon vertex by its tree-level 
counterpart. 
We find that if the same approximation is simultaneously 
implemented at the level of the SDE that governs the ghost propagator,
and the two equations are regarded as coupled, the error is $47\%$.
Instead, if the ghost dressing functions is used 
as an external input obtained from the lattice~\cite{Boucaud:2018xup,Aguilar:2021okw}, 
the error is reduced by a factor of two.

The article is organized as follows. 
In Sec.~\ref{sec:back} we introduce the relevant Green's functions 
and summarize some of their main theoretical properties.
In Sec.~\ref{sec_WTIbbcc} we derive the exact 
all-soft limit of the \bbcc vertex from the STI that it satisfies.
In Sec.~\ref{sec:SDE_deriv} we derive the 
result of the previous section at the level of the SDE that governs the \bbcc vertex.
In Sec.~\ref{sec_numerics} we use the above exact result 
in order to estimate the error induced when one of the 
ingredients of the SDE is approximated by its tree-level value. 
Then, in Sec.~\ref{sec:conc} we present our discussion and conclusions.
Finally, 
the BFM Feynman rules necessary for our calculations are listed in  Appendix~\ref{sec:App_feynman}.

\section{Theoretical background}
\label{sec:back}

In this section we introduce the notation and main theoretical elements  
needed in the present work. 

When the BFM is applied on the 
pure SU(3) Yang-Mills theory that we consider in this work, 
the gluon $A_\mu$ is split into a 
background ($B$) and a quantum ($Q$) component, according to  
$A_\mu = B_\mu + Q_\mu$. Note that only the quantum gluons 
may enter inside loops, while the background fields 
may appear only as external insertions~\cite{Abbott:1981ke}. 
The presence of these two gauge fields  
induces a considerable proliferation of Green's function, composed by combinations of $B$ and $Q$ fields~\cite{Binosi:2009qm}. 
In addition, a special gauge-fixing procedure is adopted, 
which preserves the invariance of the action under 
background gauge transformations; consequently 
the STIs triggered with respect to 
background gluons are Abelian~\cite{Abbott:1980hw}. 

The subset of BFM Green's functions composed exclusively out of quantum 
gluons corresponds precisely to those obtained 
within the linear covariant ($R_\xi$) gauges. 
In what follows we will identify the quantum gauge-fixing parameter 
$\xi_{\s Q}$ of the BFM, used to define the propagator 
$\langle 0 \vert \,T \!\left [{Q}^a_\mu(x)  {Q}^b_\nu(y) 
\right]\!\vert 0 \rangle$,
with the gauge-fixing parameter $\xi$ introduced in the renormalizable 
$R_\xi$ gauges, \ie $\xi_{\s Q} =\xi$~\cite{Aguilar:2016vin}. Thus, the full gluon propagator $\Delta^{ab}_{\mu\nu}(q) = -i\delta^{ab}\Delta_{\mu\nu}(q)$ is given by
\begin{align} 
  \Delta_{\mu\nu}(q) = P_{\mu\nu}(q)\Delta(q)  + \xi \frac{q_{\mu}q_{\nu}}{q^4} \,; \quad\quad P_{\mu\nu}(q) = g_{\mu\nu} - \frac{q_\mu q_\nu}{q^2} \,, 
\label{eq_glprop}
\end{align}
where $\Delta(q)$ denotes 
the scalar form factor of the gluon propagator. 

We emphasize that we will be working in the Landau 
gauge, corresponding to $\xi=0$.
However,  due to the particularities of the 
BFM vertices discussed below, the implementation of the 
limit $\xi\to 0$ is rather subtle, and the 
gluon propagator with a general 
value of $\xi$, as defined in \1eq{eq_glprop},
needs to be employed in intermediate steps. 

In addition, we will use extensively the full ghost propagator, \mbox{$D^{ab}(q) = i\delta^{ab} D(q)$}, 
and the corresponding dressing function, $F(q)$,
defined as
\begin{align} 
D(q) = \frac{F(q)}{q^2} \,.
\label{eq_ghprop}
\end{align}
%
\begin{figure}[t]
 \centering
\includegraphics[scale=0.5]{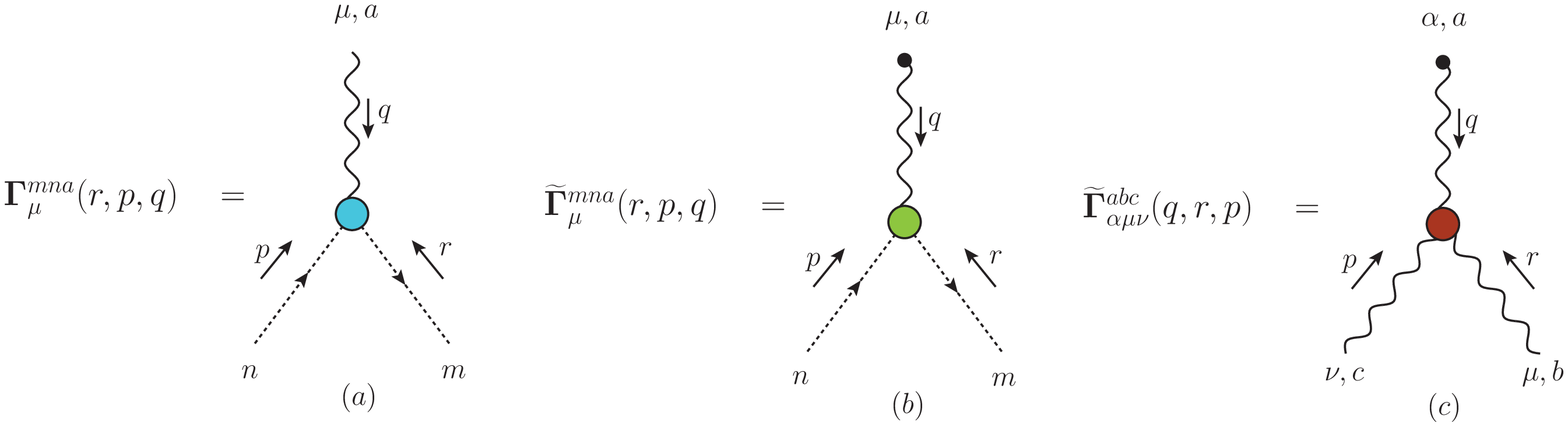}
\caption{Diagrammatic representations of the fully dressed three-point vertices.   We show in the panels $(a)$ the conventional ghost-gluon vertex ($\rm Q\cc c$),  $(b)$ the background ghost-gluon vertex (\bcc), and $(c)$ the background three-gluon vertex (\bqq),  with their respective momenta conventions.   All momenta are incoming,  \mbox{$q +r+p = 0$}.}
\label{fig_vertices3}
\end{figure}

Turning to the  three-point  sector of the theory,  in   \fig{fig_vertices3} we show the full vertices relevant for our analysis: the   conventional  ghost-gluon vertex ($\rm Q\cc c$) in panel $(a)$,   
the background ghost-gluon vertex ($\rm B\cc c$)  
in panel $(b)$, and the 
background three-gluon vertex ($\rm BQQ$) in panel $(c)$. 

Factoring out the corresponding color structures and the coupling $g$, we define the vertices $\gp$ that 
will be used for the rest of this work 
as follows\footnote{
Vertices with a single $B$-gluon  carry a ``tilde'', 
while those with more $B$-gluons carry a ``hat''.}:
\bea
\Gamma_{\bar{c}^m c^nQ_{\mu}^a}(r,p,q) &=&  -gf^{mna}\gp_{\mu}(r,p,q)\,,  \nonumber \\ 
\Gamma_{\bar{c}^m c^nB_{\mu}^a}(r,p,q) &=&-g f^{mna}\gt_\mu(r,p,q)\,,  \nonumber \\
\Gamma_{\!B_{\alpha}^a Q_{\mu}^b Q_{\nu}^c}(q,r,p)  &=& gf^{abc}\gt_{\alpha \mu \nu}(q,r,p) \,. 
\label{def3g}
\eea

We now briefly summarize some basic properties of the aforementioned vertices.   We start with  the conventional  ghost-gluon vertex,  $\gp_{\mu}(r,p,q)$, whose tensorial decomposition is  given by  
\begin{equation}  
\gp_\mu(r,p,q)=B_1(r,p,q) r_\mu + B_2(r,p,q) q_\mu \,,
 \label{eq_B12}
\end{equation}
where $B_1(r,p,q)$ and $B_2(r,p,q)$ are the corresponding form factors.  At tree-level,   \mbox{$\gp_\mu^{(0)}=r_\mu$}, and therefore \mbox{$B_1^{(0)} = 1$}, and \mbox{$B_2^{(0)} = 0$}. 
$\gp_{\mu}(r,p,q)$ satisfies the STI
\begin{equation} 
\gp_{\mu}(r,p,q)=r^\nu H_{\nu\mu}(r,p,q) \,,
 \label{eq_gammaH}
\end{equation}
where $H_{\nu\mu}(r,p,q)$ is the \emph{ghost-gluon scattering kernel}~\cite{Ball:1980ax,Aguilar:2018csq}. 

Due to Taylor's theorem~\cite{Taylor:1971ff},  in the limit of  \mbox{$p \to 0$},   
known as ``Taylor kinematics'' or ``soft ghost limit''~\cite{Fischer:2006ub,Aguilar:2009nf,Aguilar:2021okw}, 
the ghost-gluon vertex reduces to its tree-level value, \ie  \mbox{$\gp_{\mu}(r,0,-r)=r_{\mu}$}.   
Similarly, under the assumption that  
$H_{\nu\mu}(r,p,q)$ contains no poles of the type $1/r^2$, 
from \1eq{eq_gammaH} follows that 
$\gp_{\mu}(r,p,q)$ vanishes
in the soft antighost limit, \ie as \mbox{$r \to 0$}; then,  
from \1eq{eq_B12} we conclude that  $B_2(0,-q,q)=0$. 

Turning to   
$\gt_\mu(r,p,q)$, in complete analogy with 
\1eq{eq_B12} we have 
\begin{equation}  
\gt_\mu(r,p,q)=\widetilde{B}_1(r,p,q)r_\mu + \widetilde{B}_2(r,p,q)q_\mu \,;
 \label{eq_B12t}
\end{equation}
its tree-level expression \mbox{$\gt_\mu^{(0)}$} is given in Eq.~\eqref{BCC0}, such that 
\mbox{$\widetilde{B}_1^{(0)} = 2$}, and \mbox{$\widetilde{B}_2^{(0)} = 1$}. 
Note that one of the distinctive features of $\gt_\mu(r,p,q)$ is 
the linear (Abelian) STI that it satisfies  
\begin{align}
q^\mu \, \gt_\mu(r,p,q)  = D^{-1}(p) - D^{-1}(r)\,.
\label{WTI_bcc}
\end{align}

Finally, consider the 
\bqq vertex, denoted by 
$\gt_{\alpha \mu \nu}(q,r,p)$.
This vertex is a central component in SDE studies of the 
gluon propagator within the PT-BFM approach, 
and several of its main properties have been explored in the 
related literature, see, \eg~\cite{Binosi:2011wi}. 
However, for the present study of the all-soft limit 
the only relevant characteristic  of $\gt_{\alpha \mu \nu}(q,r,p)$ 
is its $\xi$-dependence at tree-level 
 [see the Feynman rule for
 $\gt^{(0)}_{\alpha \mu \nu}$ given in  Eq.~\eqref{BQQ0}]. 
 Specifically, 
 we can decompose  the full  $\gt_{\alpha \mu \nu}(q,r,p)$  as 
\begin{align} 
\gt_{\alpha \mu \nu}(q,r,p)= \widetilde{\Gamma}_{\alpha \mu \nu}(q,r,p) + \frac{1}{\xi} \left[g_{\alpha\nu} \, r_\mu - g_{\alpha\mu} \, p_\nu \right]\,,
\label{eq_bqq}
\end{align}
where the second term on the r.h.s. is the $\xi$-dependent  tree-level term.  Then, combining  Eq.~\eqref{eq_bqq} and Eq.~\eqref{BQQ0}, we see immediately that, at tree-level, \mbox{$\widetilde{\Gamma}_{\alpha \mu \nu}^{(0)}(q,r,p) = \gp^{(0)}_{\alpha \mu \nu}(q,r,p)$}, where \mbox{$\gp_{\alpha\mu\nu}^{(0)}(q,r,p) = g_{\mu \nu}(r-p)_{\alpha} + g_{\alpha \nu}(p-q)_{\mu} +  g_{\alpha \mu}(q-r)_{\nu}$}
is the standard tree-level expression of the conventional three-gluon vertex (QQQ). 

\begin{figure}[t]
\centering
\includegraphics[scale=0.5]{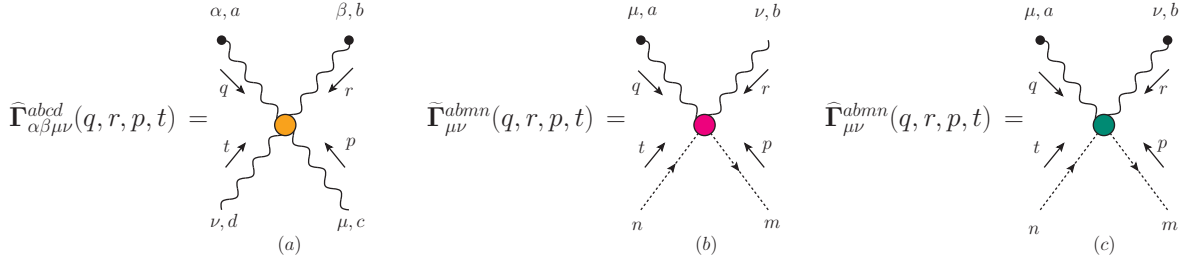}
 \caption{Diagrammatic representations of the fully dressed four-point functions:  (a) \bbqq,  (b) \bqcc,  and  (c) \bbcc.   Notice that  all  momenta are incoming,  \ie 
 \mbox{$q + r + p +t = 0$},  and  we have factored out  $-ig^2$  following the
 definitions of Eq.~\eqref{def4g}.}
\label{fig:vertices4}
\end{figure}
We now turn our attention to the four-point sector of the theory.   In \fig{fig:vertices4} we show the three four-point vertices relevant for our analysis, namely the BBQQ vertex [panel (a)], the 
\bqcc vertex [panel (b)], and the  \bbcc vertex [panel (c)].
These vertices will be denoted as 
\bea
\Gamma_{\!B_{\alpha}^a B_{\beta}^bQ_{\mu}^c Q_{\nu}^d}(q,r,p,t) &=&  -ig^2\gh^{abcd}_{\alpha\beta\mu\nu}(q,r,p,t) \,, \nonumber \\
\Gamma_{\!B_{\mu}^a Q_{\nu}^b\bar{c}^m c^n}(q,r,p,t) &=&  -ig^2{\gt}^{abmn}_{\mu\nu}(q,r,p,t)\,,  \nonumber \\ 
\Gamma_{\!B_{\mu}^a B_{\nu}^b\bar{c}^m c^n}(q,r,p,t)  &=&  -ig^2\gh^{abmn}_{\mu\nu}(q,r,p,t) \,, 
\label{def4g}
\eea
and their corresponding tree-level expressions may be found in Eqs.~\eqref{BBQQ0}, \eqref{BQcc0}, and~\eqref{BBcc0}, respectively. 

Note that $\gh^{abcd}_{\alpha\beta\mu\nu}(q,r,p,t)$ 
depends on $\xi$ already at tree level, and can be written as 
\begin{align}   
    \gh^{abcd}_{\alpha\beta\mu\nu}(q,r,p,t) = 
    \widehat{\Gamma}^{abcd}_{\alpha\beta\mu\nu}(q,r,p,t)+  \frac{1}{\xi}\left[ f^{acx}f^{xbd} g_{\alpha \mu} \, g_{ \beta \nu} - f^{adx}f^{xbc} g_{\alpha \nu} \, g_{\beta \mu}\right] \,. 
 \label{eq_bbqq}
\end{align}

 In addition, the vertex  ${\gt}^{abmn}_{\mu\nu}(q,r,p,t)$  is related to the $\gp_\nu(r,p,q)$ through the simple STI~\cite{Aguilar:2006gr,Binosi:2008qk}
\begin{align}   
q^\mu {\gt}^{abmn}_{\mu\nu} (q,r,p,t) \! = \!f^{nax}f^{bmx}\gp_\nu(p,q\!+\!t,r) +f^{nbx}f^{max} \gp_\nu (q\!+\!p,t,r) +f^{nmx}f^{abx}\gp_\nu(p,t,q\!+\!r) \,. 
\label{eq_WTIBQcc}
\end{align}

The vertex $\gh^{abmn}_{\mu\nu}(q,r,p,t)$, which  is central to our analysis, may be expanded as~\cite{Pascual:1980yu,Binosi:2014kka,Huber:2017txg}
\begin{equation}    
    \gh^{abmn}_{\mu \nu}(q,r,p,t) =  \sum_{i=1}^{10}\sum_{j=1}^{8}{T}_{ij} (q,r,p,t)  {\ell}^i_{\mu \nu} c^{abmn}_j \,,
\label{eq_npbbcc}
\end{equation}
where 
\begin{equation}
    \begin{tabular}{lllll}
        $\ell^1_{\mu \nu}  =  g_{\mu \nu}  \,,$   \hspace{0.6 cm} &
        $\ell^2_{\mu \nu}  =   q_\mu r_\nu \,,$  \hspace{0.6cm}  &
        $\ell^3_{\mu \nu}  = q_\mu p_\nu \,,$   \hspace{0.6cm} &
        $\ell^4_{\mu \nu}  =   q_\nu r_\mu \,,$  
        \hspace{0.6cm} &
        $\ell^5_{\mu \nu}  =  q_\nu p_\mu \,,$ \\
        $\ell^6_{\mu \nu}  = r_\mu p_\nu \,,$  &
        $\ell^7_{\mu \nu} =  p_\mu r_\nu  \,,$ &
        $\ell^8_{\mu \nu} =  q_\mu q_\nu \,, $ &
        $\ell^9_{\mu \nu}  =  r_\mu r_\nu \,, $  &
        $\ell^{10}_{\mu \nu} = p_\mu p_\nu \,, $
    \end{tabular}
\label{eq_lorentzbasis}    
\end{equation}
and
\begin{equation}
\hspace{-0.2cm}
    \begin{tabular}{llll}
$c_1^{abmn}  =  f^{anx}f^{mbx}  \,,$  \hspace{0.2cm} &   
$c_2^{abmn}=f^{max}f^{bnx} \,,$
 \hspace{0.2cm} & 
$c_3^{abmn}  = \delta^{ab} \delta^{mn} \,,$
 \hspace{0.2cm} & 
 $c_4^{abmn} = \delta^{am} \delta^{nb} \,,$  \\
  $c_5^{abmn}  =  \delta^{an} \delta^{bm} \,,$ &   $c_6^{abmn}  = d^{abr}f^{mnr} \,,$  &
 $c_7^{abmn} =  d^{amr}f^{bnr} \,,$     &  $c_8^{abmn} = d^{anr}f^{bmr}  \,.$ 
\end{tabular}
\label{eq_colorbasis}
\end{equation}
At tree level, only 
\mbox{$T^{(0)}_{11}= T^{(0)}_{12}= 1$} 
are nonvanishing. 

The Bose symmetry of $\gh^{abmn}_{\mu \nu}(q,r,p,t)$ under 
the exchange of two background gluons, 
\ie $(a,\mu,q) \leftrightarrow (b,\nu,r)$, imposes 
additional constraints on the form factors ${T}_{ij} (q,r,p,t)$. Specifically, 45 out of the 80 form factors $T_{ij}(q,r,p,t)$ can be written as permutations of the arguments of the remaining 35, \eg $T_{73}(q,r,p,t)=T_{33}(r,q,p,t)$.

Of course, in the all-soft limit that we study, the 
tensorial structures collapse to $g_{\mu \nu}$, 
which, by virtue of the Bose symmetry, may be 
multiplied by 
\mbox{$(f^{anx}f^{mbx}+f^{max}f^{bnx})$}, 
\mbox{$(\delta^{am} \delta^{nb}+\delta^{an}\delta^{bm})$}, and 
\mbox{$\delta^{ab} \delta^{mn}$}. However, only the first 
color combination respects the ghost-antighost symmetry of the vertex, so that 
we finally arrive at the unique structure relevant for the 
all-soft limit, namely \mbox{$(f^{anx}f^{mbx}+f^{max}f^{bnx})g_{\mu \nu}$}.


Let us finally introduce the renormalization constants $Z_i$ that 
connect bare and renormalized quantities.
In particular, we have~\cite{Aguilar:2013xqa,Aguilar:2016vin} 
\begin{align}
\label{renor_prop}
\Delta_{\s R}(q^2)= Z^{-1}_{A} \Delta(q^2),\qquad  F_{\s R}(q^2)= Z^{-1}_{c} F(q^2), \qquad  g_{\s R} = Z_g^{-1} g \,, 
\end{align}
and 
\bea 
\label{eq_z14}
    \gp_\mu(r,p,q) &:= Z_1^{-1} \gp^\srm{R}_\mu(r,p,q) \,, \qquad \qquad \gt_{\mu\nu}(q,r,p,t) := \zt^{-1}_4 \gt^\srm{R}_{\mu\nu}(q,r,p,t) \,, \nonumber \\
    \gt_\mu(r,p,q) &:= \zt^{-1}_1 \gt^\srm{R}_\mu(r,p,q) \,, \qquad \qquad \gh_{\mu\nu}(q,r,p,t) := \widehat{Z}^{-1}_4 \gh^\srm{R}_{\mu\nu}(q,r,p,t) \,,
\eea
where we have omitted the color structures for simplicity.

By virtue of the various STIs relating the above Green's functions, 
the renormalization constants satisfy the conditions  
\begin{align}
\widehat{Z}_4=\zt_1\,, \qquad  \zt_1=Z_c  \,, \qquad \zt_4=Z_1\,, \, \qquad  Z^{-1}_g = Z_1^{-1} Z_A^{1/2} Z_c \,.
\label{STIcon}
\end{align}
Note that, in the Landau gauge, the renormalization 
constant $Z_1$ is 
finite (cutoff-independent), as a consequence of 
Taylor's theorem~\cite{Taylor:1971ff}.

\section{All-soft limit: an exact result}   
\label{sec_WTIbbcc}
  
  In this section we derive the exact 
  all-soft limit of the vertex ${\gh}^{abmn}_{\mu\nu} (q,r,p,t)$, 
  by resorting to the 
simple STI that relates ${\gh}^{abmn}_{\mu\nu} (q,r,p,t)$ 
with the $\gt_\nu$ of Eq.~\eqref{eq_B12t},  namely
\begin{align} \label{eq_WTIBBcc}
    q^\mu \gh_{\mu\nu}^{abmn}(q,r,p,t)\! = \! f^{abx}f^{mnx}\gt_\nu(p,t,q+r) + f^{anx}f^{bmx}\gt_\nu(p,q+t,r)+ f^{amx}f^{nbx}\gt_\nu(q + p, t, r)  \,.
\end{align}
\1eq{eq_WTIBBcc}
may be derived within the systematic approach  
provided by the Batalin-Vilkovisky formalism~\cite{Batalin:1977pb,Batalin:1983ggl,Binosi:2002ez,Binosi:2009qm}. 
A more direct derivation 
relies on the observation that, at tree-level
[see \2eqs{BCC0}{BBcc0}], if we contract $\gh^{(0)abmn}_{\mu\nu}$ 
by $q^\mu$ and, with the aid of the Jacobi identity,
add to the answer
$(p-t)_{\nu}[f^{abx}f^{mnx} +  f^{anx}f^{bmx} + f^{amx}f^{nbx}]=0$, 
we have 
\be
q^\mu  \gh^{(0)abmn}_{\mu\nu} = f^{abx}f^{mnx} \underbrace{(p-t)_{\nu}}_{\gt^{(0)}_\nu(p,t,q+r)} + 
f^{anx}f^{bmx} \underbrace{(p-t-q)_{\nu}}_{\gt^{(0)}_\nu(p,q+t,r)}+ f^{amx}f^{nbx} \underbrace{(p-t+q)_{\nu}}_{\gt^{(0)}_\nu(q + p, t, r)} \,.
\label{WTIBBcctree}
\ee
Since the STIs triggered with respect  
to a background leg 
maintain their tree-level form, the naive generalization of 
\1eq{WTIBBcctree} leads us to \1eq{eq_WTIBBcc}. 

The way to proceed 
with \1eq{eq_WTIBBcc}
is analogous to the typical derivation of 
a WI out of a STI: 
as the momentum that triggers the STI tends to zero, 
a Taylor expansion of both sides is carried out, followed 
by an appropriate matching of terms linear in $q$.
In the case of the  four-particle vertex 
$\gh_{\mu\nu}^{abmn}(q,r,p,t)$
that we consider, 
it is convenient to choose as a point of departure 
the special kinematic configuration 
\mbox{$(q,r,p,t) \to (q, -q, 0,0)$}.
In that case, \1eq{eq_WTIBBcc} reduces to 
\begin{align} \label{eq_WTIbbcc0}
    q^\mu \gh_{\mu\nu}^{abmn}(q,-q,0,0)\! = \! f^{abx}f^{mnx}\gt_\nu(0,0,0) + f^{anx}f^{bmx}\gt_\nu(0,q,-q)+ f^{amx}f^{nbx}\gt_\nu(q, 0, -q)  \,.
\end{align}

To begin with, it is clear that that \mbox{$\gt_\nu(0,0,0) = 0$},  since it has just one Lorentz index and all momenta were set to zero. 

Then,  the second and third terms on the r.h.s of Eq.~\eqref{eq_WTIbbcc0} correspond to the ``soft  antighost'' (\ie~\mbox{$r\to 0$} ) 
 and ``soft ghost'' (or equivalently ``Taylor kinematics'' where \mbox{$p\to 0$}) limits of the background ghost-gluon vertex, $\gt_\mu$, respectively.   
 
 The derivation of the special exact relation that $\gt_\mu$,  in the  ``soft ghost'' kinematics satisfies, was shown in~\cite{Aguilar:2017dco} using three different approaches.  In what follows,  we will sketch some of the main steps of the derivation based on the STI that $\gt_\mu$ satisfies, since we are using a different tensorial basis for $\gt_\mu$. These steps  will be also relevant for the derivation of the ``soft antighost'' limit.
 
The starting point is the combination of the most general tensorial decomposition of $\gt_\mu$,  written in Eq.~\eqref{eq_B12t},  and the STI of Eq.~\eqref{WTI_bcc} that $\gt_\mu$ satisfies, which lead us to  
\be 
( q\cdot r) \widetilde{B}_1(r,p,q) + q^2 \widetilde{B}_2(r,p,q) = D^{-1}(p) - D^{-1}(r) \,. \label{WTI_bcc_ffs}
\ee 

Next, assuming that there are no poles associated to the \mbox{$r = 0$} and \mbox{$p = 0$} limits; then, in the soft ghost limit (\mbox{$p = 0$} and \mbox{$r = - q$})~\1eq{WTI_bcc_ffs} becomes 
\be 
\widetilde{B}_1(q,0,-q) - \widetilde{B}_2(q,0,-q) = F^{-1}(q) \,. 
\label{Bcc_soft_ghost_ffs}
\ee 
Thus, setting $p=0$ and  $r = - q$ in Eq.~\eqref{eq_B12t} and in the sequence, using the Eq.~\eqref{Bcc_soft_ghost_ffs}, we find that in the soft ghost limit 
\be 
\gt_\mu(q,0,-q) = q_\mu\left[ \widetilde{B}_1(q,0,-q) - \widetilde{B}_2(q,0,-q) \right] = q_\mu F^{-1}(q) \,. 
\ee
 Similarly, in the soft antighost limit (\mbox{$r = 0$} and \mbox{$p = - q$}), \1eq{WTI_bcc_ffs} simplifies to  
\be 
\widetilde{B}_2(0,q,-q) = F^{-1}(q) \,, 
\label{Bcc_soft_antighost_ffs}
\ee 
where we used Lorentz invariance to change the sign of the  arguments of the scalar function $\widetilde{B}_2$. Then, substituting the above result in \1eq{eq_B12t}, we obtain the exact relation 
\be 
\gt_\mu(0,q,-q) = - q_\mu \widetilde{B}_2(0,q,-q) = - q_\mu F^{-1}(q) \,. \label{Bcc_soft_antighost}
\ee 

Therefore, we find that both limits are related to each other as 
\begin{align}
\gt_\mu(q,0,-q) = - \gt_\mu(0,q,-q) = q_\mu F^{-1}(q) \,.
\label{limit_back} 
\end{align}

Let us mention in passing that the results of  
\1eq{limit_back}, derived above in full generality,  
may also be obtained from the standard gauge technique Ansatz~\cite{Ball:1980ax}, 
\begin{align}   
    \gt_\mu(r,p,q)=\left[\frac{D^{-1}(p)-D^{-1}(r)}{p^2-r^2} \right] (2r+q)_{\mu}  
 + {\mathcal A}_{\rm T}(r,p,q)[(r\cdot q)  q_{\mu} -q^2 r_{\mu}]  \,, 
\label{eq_BCbcc}
\end{align}
provided that the undetermined transverse 
(automatically conserved) part 
is well-behaved in the limit $q \to 0$.

Substituting the results of \1eq{limit_back} into the r.h.s. of 
\1eq{eq_WTIbbcc0}, we arrive at 
\begin{align}
q^\mu \gh^{abmn}_{\mu\nu}(q,0,-q,0)
&=q_\nu(f^{max}f^{xbn} + f^{mbx}f^{xan})F^{-1}(q) \,,
\end{align}
which, upon expansion around $q=0$, yields 
the final exact result 
\begin{align}
   \gh^{abmn}_{\mu\nu}(0,0,0,0)=g_{\mu\nu}(f^{max}f^{bnx} + f^{mbx}f^{anx})F^{-1}(0) \,.
  \label{eq_bbccall}
\end{align}
Clearly, at tree-level, when \mbox{$F^{-1}(0)=1$}, the above result 
reduces simply to the momentum-independent 
expression for $\gh^{(0)abmn}_{\mu\nu}$ given in Eq.~\eqref{BBcc0}.

Finally, in terms of the form factors ${T}_{ij}(q,r,p,t)$ appearing in \1eq{eq_npbbcc}, 
the exact result of \1eq{eq_bbccall} implies that 
\be
T(0):= T_{11}(0,0,0,0)=T_{12}(0,0,0,0) = F^{-1}(0) \,.
\label{T0}
\ee
The remaining form factors ${T}_{ij}(0,0,0,0)$ are undetermined, 
because their associated tensor structures vanish directly in the all-soft limit.

\section{All-soft limit of the SDE}
\label{sec:SDE_deriv}

In this section we derive the all-soft limit 
of the vertex $\gh^{abmn}_{\mu\nu}(q,r,p,t)$ 
from the SDE that it satisfies. This is a subtle 
exercise, mainly for two reasons: first, a series of key   
gauge cancellations must be implemented before 
the Landau gauge limit may be taken safely; and second, 
several instrumental properties of the 
vertices nested inside the Feynman diagrams 
of the SDE must be employed, in order for the result 
of \1eq{eq_bbccall} to emerge.

\begin{figure}[t]
\centering 
\includegraphics[scale=0.5]{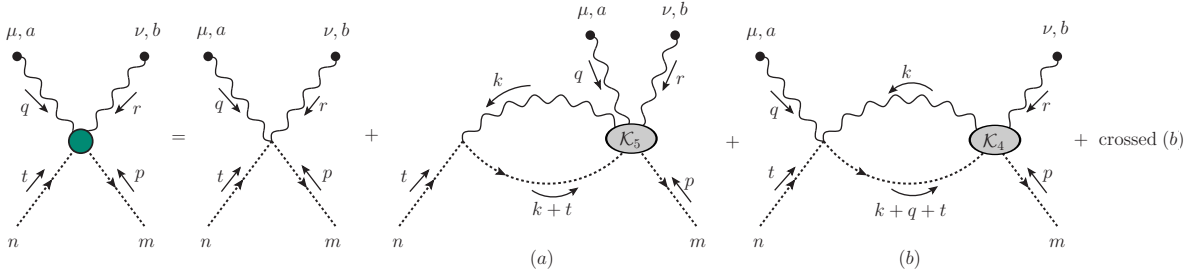}
 \caption{Diagrammatic representation of the SDE for the \bbcc.  The gray ellipses represent the five- and four-points kernels, ${\cal K}^{abcem}_{5\,\mu\nu\rho}$ and ${\cal K}^{bcem}_{4\,\mu\nu}$, respectively. The diagram obtained from $(b)$ through  crossing of the gluon legs,  \mbox{ $\mu,a  \leftrightarrow \nu,b$}, 
 is not shown.}
\label{fig:compact}
\end{figure}

We find it convenient to set up the SDE 
with respect to the ghost field carrying the momentum $t$. 
According to the standard procedure, the ghost field is attached to all possible tree-level vertices containing it, while the remaining three fields connect to the diagram by means of appropriate dressed kernels. 
In the particular case 
that we consider there are two relevant tree-level vertices: 
the standard \acc vertex, 
and the  \bqcc vertex that is particular to the BFM. 
The resulting SDE is represented 
diagrammatically in  Fig.~\ref{fig:compact},
in terms of the five- and four-particle kernels, 
denoted by ${\cal K}^{abcme}_{5\,\mu\nu\rho}$ and  
${\cal K}^{bcme}_{4\,\nu\rho}$, respectively; 
the crossed diagram obtained by interchanging the background gluons of $(b)$ is not shown. 

\begin{figure}[h]
\centering
\includegraphics[width=0.98\textwidth]{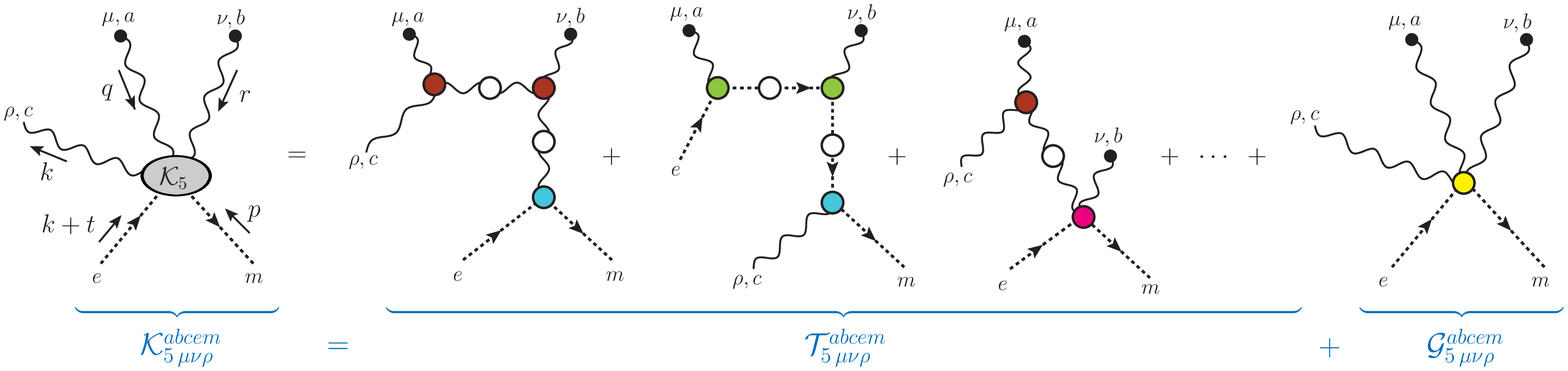}
\includegraphics[width=0.65\textwidth]{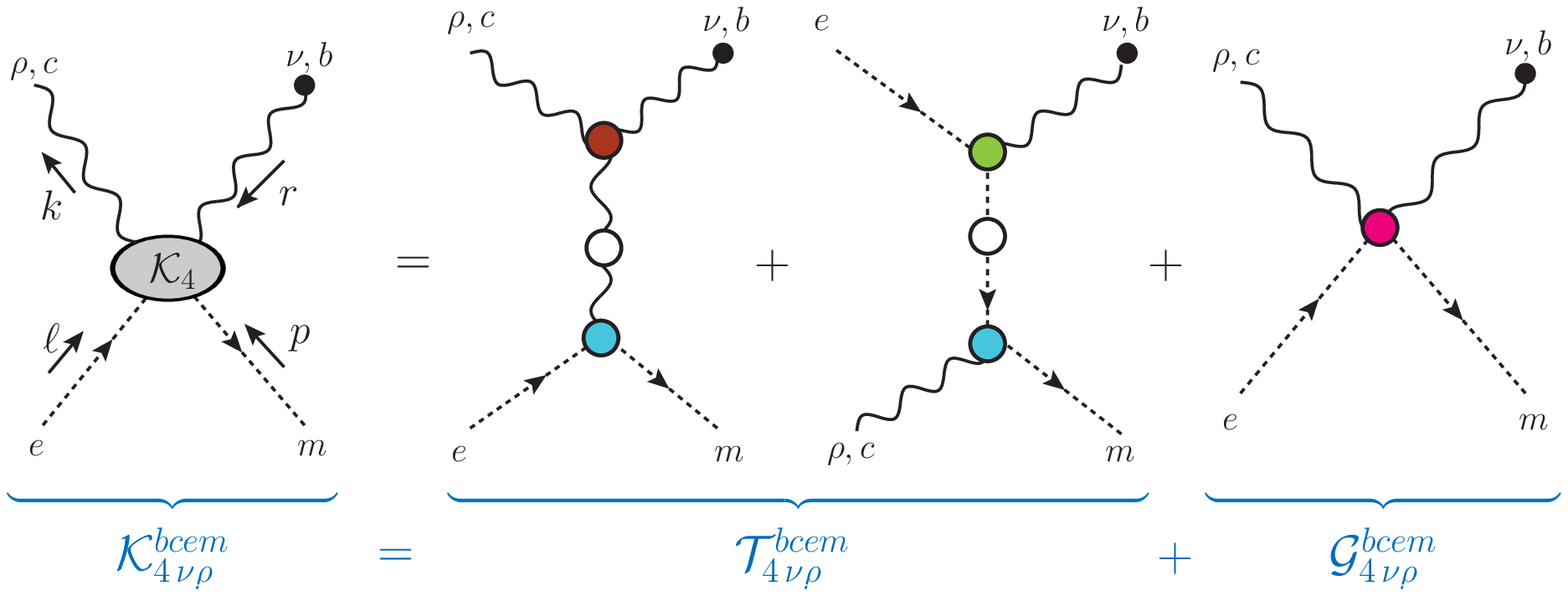}
 \caption{{\it First row}: The skeleton expansion of the five-particle kernel ${\cal K}^{abcem}_{5\,\mu\nu\rho}$ is shown as the 
 sum of the 1PR and 1PI contributions, denoted by  
 ${\cal T}^{abcem}_{5\;\mu\nu\rho}$ and 
 ${\cal G}^{abcem}_{5\,\mu\nu\rho}$, respectively; the dots  
 indicate additional 1PR graphs not shown. 
 {\it Second row}:
 The corresponding contributions of 
 ${\cal K}^{bcem}_{4\,\nu\rho}$, denoted by  
 ${\cal T}^{bcem}_{4\;\nu\rho}$ and ${\cal G}^{bcem}_{4\,\nu\rho}$, respectively.} 
\label{fig_kernel5}
\end{figure}

Each of the kernels ${\cal K}^{abcme}_{5\,\mu\nu\rho}$ and 
${\cal K}^{bcme}_{4\,\nu\rho}$ consists of 
a component that sums up the one-particle reducible (1PR) terms, 
to be denoted by ${\cal T}^{abcme}_{5\;\mu\nu\rho}$ 
and ${\cal T}^{bcme}_{4\;\nu\rho}$, 
and a component  
containing all possible one-particle irreducible (1PI) contributions, 
to be denoted by ${\cal G}^{abcme}_{5\;\mu\nu\rho}$ and 
${\cal G}^{bcme}_{4\;\nu\rho}$, respectively, as shown in Fig.~\ref{fig_kernel5}. 
Thus, we have 
\be
{\cal K}^{abcme}_{5\;\mu\nu\rho} = 
{\cal T}^{abcme}_{5\;\mu\nu\rho} + \,
{\cal G}^{abcme}_{5\;\mu\nu\rho} \,, \qquad \qquad 
{\cal K}^{bcme}_{4\;\nu\rho} = 
{\cal T}^{bcme}_{4\;\nu\rho} + \,
{\cal G}^{bcme}_{4\;\nu\rho} \,.
\ee
Note that 
${\cal G}^{abcme}_{5\;\mu\nu\rho}$ 
coincides with 
the dressed loop-wise (skeleton)  
expansion of the 1PI five-point Green's function 
$\langle 0 \vert \,T \!\left [{B}^a_\mu \, {B}^b_\nu \,{Q}^c_\rho \, {\bar c}^{m} \,c^{e} \right]\!\vert 0 \rangle$, 
while ${\cal G}^{bcme}_{4\;\nu\rho}$ corresponds to the 1PI 
four-point function
$\langle 0 \vert \,T \!\left [{B}^b_\nu\,
{Q}^c_\rho \, {\bar c}^{m} \, c^{e} \right]\!\vert 0 \rangle$.

\begin{figure}[t]
 \centering
 \includegraphics[width=0.98\textwidth]{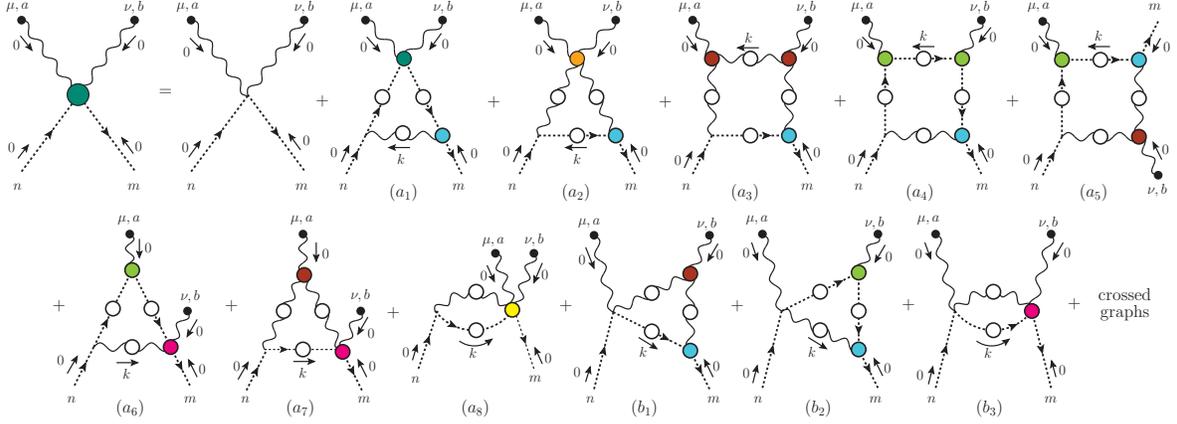}
 \caption{The diagrammatic representation of the SDE for the vertex \bbcc,  expanded in terms of 1PI vertices; crossed diagrams are not shown. Note that diagrams $(a_i)$ with $i=1,2,\cdots,8$ originate from the expansion of the 1PR kernel in $(a)$ in \fig{fig:compact}, while $(b_i)$ with $i=1,2,3$ come from $(b)$.}
\label{fig_completed}
\end{figure}

 The renormalization of the SDE for the vertex $\gh^{abmn}_{\mu\nu}(q,r,p,t)$ proceeds by introducing the renormalization relations given in 
\2eqs{renor_prop}{eq_z14} into each 
 of the graphs in \fig{fig_completed}, employing 
 the constraints listed in \1eq{STIcon}. 
Then, the renormalized version of the vertex SDE reads
 \begin{align}
 \label{SDE_bbcc}
 \gh^{abmn}_{\mu\nu}(q,r,p,t) =  Z_c\,\gh^{(0)abmn}_{\mu\nu}  -ig^2 Z_1    \sum  {\rm all\,\,graphs}\,,
 \end{align}
where all subscripts ``\rm{R}''  have been suppressed in order 
to avoid notation clutter.

We next evaluate the all-soft limit 
of the diagrams given in \fig{fig_completed}.
In doing so, the cancellation of terms proportional to $1/\xi$ 
must be carried out before the Landau limit, $\xi\to 0$, is taken.

({\it i})
We start by noticing that the diagrams 
$(a_1)$, $(a_4)$, $(a_6)$, and $(b_2)$ do not 
contain vertices with $\xi$-dependent tree-level expressions;  
therefore, the Landau gauge may be reached directly  
by setting $\xi=0$ throughout.
Then, it is elementary to establish that $(a_1)$, $(a_4)$, and $(a_6)$
vanish in the all-soft limit, $(p,q,r,t)=(0,0,0,0)$, 
because the sequence 
\be
(p+k)^{\sigma}\Delta_{\sigma\rho}(k) \xrightarrow[\text{}]{\text{$\xi=0$}}
(p+k)^{\sigma} P_{\sigma\rho}(k)\Delta(k)
\xrightarrow[\text{}]{\text{$p=0$}}
k^{\sigma} P_{\sigma\rho}(k)\Delta(k) =0 \,,
\label{sec}
\ee
is triggered. As for graph $(b_2)$, it vanishes because it 
contains the ghost-gluon vertex in the soft antighost limit, 
\ie $\gp_\rho(0,k,-k)=0$. Thus, in total,  
\begin{equation}
    (a_1)^{abmn}_{\mu\nu} = (a_4)^{abmn}_{\mu\nu} = (a_6)^{abmn}_{\mu\nu} =  (b_2)^{abmn}_{\mu\nu} = 0 \,. 
\end{equation}

({\it ii})
Diagrams $(a_2)$, $(a_3)$, $(a_5)$, and $(b_1)$ 
eventually vanish, because they all contain the ghost-gluon vertex in the soft antighost limit.
However, since these graphs
contain tree-level 
vertices with terms proportional to $1/\xi$, a limiting procedure must be 
followed in order to safely implement the Landau gauge, 
thus triggering the result $\gp_\rho(0,k,-k)=0$.
To that end, in the case of graphs  $(a_2)$, $(a_3)$, and $(a_5)$  
note that we ``gain'' a power of $\xi$ by employing   
$(p+k)^{\sigma}\Delta_{\sigma\rho}(k) 
\xrightarrow[\text{}]{\text{$p=0$}}
k^{\sigma} \Delta_{\sigma\rho}(k)=  \xi \,k_{\rho}/k^2$.
The presence of this $\xi$ makes all $\xi$-independent terms vanish, 
as $\xi\to 0$. Furthermore,  it 
cancels the $1/\xi$ terms originating from the vertices, 
furnishing finite expressions, given by 
\bea  
(a_{2})^{abmn}_{\mu\nu} & \! = & \! g^4c_1^{cndm} \!\! \int_k \!\! \left(\frac{k^\rho}{k^2}\right) D(k)  \Delta^{\sigma \lambda}(k) \gp_\lambda(0,-k,k)  \left[c_1^{adbc} g_{\mu \sigma} g_{\nu \rho} - c_1^{acbd} g_{\mu \rho} g_{\nu \sigma}\right]\,, \nonumber \\
(a_3)_{\mu\nu}^{abmn} &=& g^4 c_1^{cdba}c_1^{dcnm} \int_k \left(\frac{k^\gamma}{k^2}\right) D(k) \gp_\rho(0,k,-k) [g_{\mu\gamma}k_\beta + g_{\mu\beta}k_\gamma] \times \nonumber  \\
 && \hspace{-2.5cm}  \qquad \qquad \qquad \left[ \Delta^{\rho\sigma}(k) \Delta^{\alpha\beta}(k)\widetilde{\Gamma}_{\nu\sigma\alpha}(0,k,-k) \,+  \left(\frac{k^\beta}{k^2}\right) \Delta^\rho_\nu(k) + \left( \frac{k^\rho}{k^2}\right)\Delta^\beta_\nu(k) \right]  \,, \label{d13eq}  \\ 
 (a_5)_{\mu\nu}^{abmn}&=&g^4c_1^{cdma}c_1^{ndbc}\int_k   \left( \frac{ k_\rho }{k^2} \right)\Delta_{\alpha\beta}(k) D^2(k)\gt_\nu(-k,k,0) \gp_\beta(0,k,-k) (g_{\mu\rho}k_\alpha + g_{\mu\alpha}k_\rho)  \,.  \nonumber
\eea

The vertex $\widetilde{\Gamma}_{\nu\sigma\alpha}(0,k,-k)$, appearing in the second equation, has been defined in Eq.~\eqref{eq_bqq}, whereas the  factor \hbox{$c_1^{cndm}$} is given in \1eq{eq_colorbasis}. In addition, we have introduced the integral measure 
\begin{equation}
\int_k := \frac{1}{(2\pi)^4} \int\!\! \dd[4]{k} \,,
\label{measure}
\end{equation}
where the use of a symmetry-preserving regularization scheme is implicitly understood. 

Since at this point we may set $\xi=0$ in the expressions of \1eq{d13eq}, 
the result $\gp_\rho(0,k,-k)=0$ makes them all vanish. 

Turning to $(b_1)$, the term $\gp_{\nu\rho\sigma}^{(0)}(0,k,-k)$ 
of the BFM tree-level vertex $\gt_{\nu\rho\sigma}^{(0)}(0,k,-k)$ 
simply yields the standard Landau gauge result, denoted by $(b_{11})$,
while the term $\xi^{-1} (g_{\nu\sigma} k_\rho + g_{\nu\rho} k_\sigma)$,
once contracted with the adjacent gluon propagators, yields a 
finite contribution, denoted by $(b_{12})$, given by 
\be
(b_{12})_{\mu\nu}^{abmn}  = g^4 c_1^{gcmb}c_1^{cnga} g_{\mu\beta} \int_k D(k) \gp_\rho(0,k,-k) \left[\left(\frac{k^\beta}{k^2}\right)  \Delta^\rho_\nu(k) + \left(\frac{k^\rho}{k^2}\right) \Delta^\beta_\nu(k)\right] \,.
\label{b12}
\ee
Since both $(b_{11})$ and $(b_{12})$ contain the vertex $\gp_\rho(0,k,-k)=0$, 
they both vanish when $\xi=0$. 
Thus, we finally have 
\begin{equation}
    (a_2)^{abmn}_{\mu\nu} = (a_3)^{abmn}_{\mu\nu} = (a_5)^{abmn}_{\mu\nu} = (b_1)^{abmn}_{\mu\nu} = 0 \,.
\end{equation}
%

({\it iii}) The treatment of diagram $(a_8)$, 
which contains all 1PI corrections of the five-particle kernel, 
denoted by ${\cal G}^{abcme}_{5\,\mu\nu\rho}$, requires particular care.  
In what follows we explain why, in the all-soft limit, $(a_8)=0$.

Since ${\cal G}^{abcme}_{5\,\mu\nu\rho}$ is computed using the 
BFM Feynman rules of Table~\ref{fig_feyback}, it is clear that 
individual diagrams may contain contributions proportional to $1/\xi$. 
Nonetheless, certain powerful formal properties 
guarantee that, due to massive cancellations among different diagrams, 
the entire ${\cal G}^{abcme}_{5\,\mu\nu\rho}$ 
contains no such terms, and therefore, the Landau-gauge 
limit may be safely implemented. 

The rather technical demonstration of the above statement 
proceeds by appealing to 
the special relations known as Background-Quantum identities 
(BQIs)~\cite{Grassi:2001zz,Binosi:2002ez,Binosi:2009qm}, 
derived through appropriate functional differentiation of the 
STI functional within the 
Batalin-Vilkovisky quantization formalism~\cite{Batalin:1977pb,Batalin:1983ggl}.
In particular, the BQIs express Green's functions 
containing background fields ($B$) in terms of ({\it i}) 
conventional Green's functions containing quantum fields ($Q$) 
and ({\it ii})  
auxiliary Green's functions involving 
the so-called ``antifields'' and 
``background sources'', arising from 
interaction terms 
particular to the aforementioned formalism.
The special Feynman rules describing these latter Green's functions may be found in Figs.~(B.3)-(B.4) of~\cite{Binosi:2009qm}; their most 
relevant feature for our purposes is that they are 
$\xi$-independent.

Note that the exact form of the BQI 
that ${\cal G}^{abcme}_{5\,\mu\nu\rho}$ satisfies is not required 
for the argument that we present; it suffices to 
know that the BQI relates 
${\cal G}^{abcme}_{5\,\mu\nu\rho}$ to 
a finite set of Green's functions, all of which 
are regular as $\xi\to 0$. Consequently, 
the limit  $\xi\to 0$ of 
${\cal G}^{abcme}_{5\,\mu\nu\rho}$ is completely well-defined. 
The immediate upshot of the above result is that 
the all-soft limit of diagram $(a_8)$ vanishes, 
simply because the Landau gauge may be 
taken directly, thus triggering the 
sequence given in \1eq{sec}.

({\it iv})
We finally arrive at the contributions that 
survive the all-soft limit: they originate from 
diagrams $(a_{7})$ and $(b_3)$,
together with their crossed counterparts, 
to be denoted by $(a_{7}^c)$ and $(b_3^c)$, respectively.

Both $(a_{7})$ and $(b_3)$ contain 
${\cal G}^{abcme}_{4\,\nu\rho}$, namely the 1PI Green's function 
$\langle 0 \vert \,T \!\left [{B}^b_\nu\,
{Q}^c_\rho \, {\bar c}^{m} \, c^{e} \right]\!\vert 0 \rangle$.
The arguments presented in ({\it iii}) for ${\cal G}^{abcme}_{5\,\mu\nu\rho}$ apply unaltered in the case of  ${\cal G}^{abcme}_{4\,\nu\rho}$, 
and the Landau gauge limit may be taken directly in it. 
Since the product   
$\gt_{\mu\rho'\sigma'}^{(0)}(0,k,-k) 
\Delta_{\rho'\rho}(k)\Delta_{\sigma'\sigma}(k)$ 
is finite as $\xi \to 0$, diagram $(a_{7})$ 
is nonvanishing, and the same is true for $(a_{7}^c)$. 
Similarly, in $(b_3)$ the ${\cal G}^{abcme}_{4\,\nu\rho}$ is connected 
to the rest of the diagram with propagators and vertices that are regular and nonvanishing as
$\xi \to 0$, and the same happens with $(b_3^c)$. 
The final individual contributions are given by 
\begin{align}   
\label{d4_bbcc}
(a_{7})_{\mu\nu}^{abmn} &= f^{ebx}f^{dnx}\int_k\Delta(k) D(k) \,P_{\nu}^{\alpha}(k) \,\gt^{adme}_{\mu\alpha}(0,-k,0,k) \,,   \nonumber \\ %
(b_3)_{\mu\nu}^{abmn} &= f^{eax}f^{dnx} \int_k \Delta(k) D(k) \,P_{\mu}^{\alpha}(k) \,\gt^{bdme}_{\nu\alpha}(0,-k,0,k) \,,   \nonumber \\ %
(a_{7}^c)_{\mu\nu}^{abmn} &= f^{enx}f^{bdx} \int_k \Delta(k) D(k) \,P_{\nu}^{\alpha}(k) \,\gt^{adme}_{\mu\alpha}(0,-k,0,k) \,,  \nonumber \\  %
(b_3^c)_{\mu\nu}^{abmn} &= f^{enx}f^{adx} \int_k \Delta(k) D(k) \,P_{\mu}^{\alpha}(k) \,\gt^{bdme}_{\nu\alpha}(0,-k,0,k) \,.
\end{align}

As a consequence of the considerations presented in ({\it i})-({\it iv}), the all-soft limit of   
\1eq{SDE_bbcc} is given by  
\begin{align}
 \label{eq_final}
 \gh^{abmn}_{\mu\nu}(0,0,0,0) = Z_c\gh^{(0)abmn}_{\mu\nu}  -ig^2Z_1\left[ (a_{7}) + (a_{7}^c)  +  (b_3) + (b_3^c) \right]_{\mu\nu}^{abmn}\,.
\end{align}
\begin{figure}[t]
    \centering
    \includegraphics[scale=0.5]{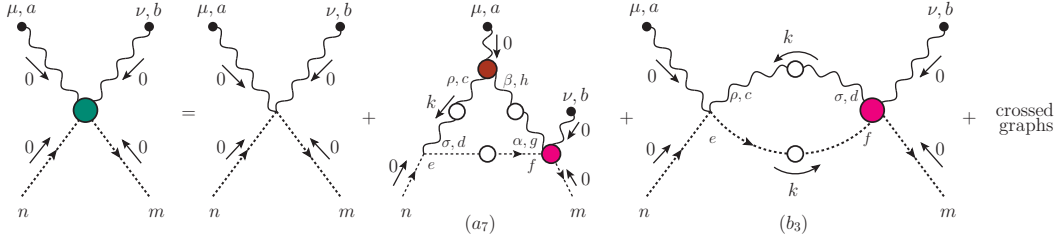}
    \caption{The diagrammatic representation of the SDE for the \bbcc in the all-soft limit.}
    \label{fig_soft}
\end{figure}

The next step consists in determining 
$\gt_{\mu\alpha}^{adme}(0,-k,0,k)$, the common ingredient 
in all integrals appearing in \1eq{d4_bbcc}.
To that end, consider the STI in \eqref{eq_WTIBQcc}, 
and implement the special kinematic configuration 
$(q,r,p,t) \to (q,-k,-q,k)$, such that 
\bea
q^\mu \gt^{adme}_{\mu\alpha}(q,-k,-q,k) &=&  f^{eax}f^{dmx} \gp_{\alpha}(-q,q+k,-k)  +  f^{edx}f^{max} \gp_{\alpha}(0,k,-k) \nonumber\\
&+& f^{emx}f^{adx} \gp_{\alpha}(-q,k,q-k) \,.
\label{stisk} 
\eea
We then carry out a 
Taylor expansion of both sides around $q=0$. It is elementary 
to show that the zeroth order in $q$ vanishes on the r.h.s. because it 
is proportional to the Jacobi identity, while the 
coefficients of the linear terms are related by   
\bea
\gt^{adme}_{\mu\alpha}(0,-k,0,k) &=&f^{aex}f^{dmx}   
\left\{ \frac{\partial}{\partial q^\mu}\left[B_1(-q,q+k,-k)q_\alpha + B_2(-q,q+k,-k)k_\alpha \right] \right\}_{q=0} \nonumber \\ 
  &&\hspace{-1.5cm}  +
    f^{mex}f^{adx} \left\{ \frac{\partial}{\partial q^\mu}  \left[B_1(-q,k,q-k)q_\alpha +B_2(-q,k,q-k)(q-k)_\alpha \right]\right\}_{q=0}\,,
\eea
where we have used that the momentum $k$ is independent of $q$. 
Then, since \mbox{$B_2(0,k,-k)=0$} [see discussion after \1eq{eq_gammaH}], we arrive at 
\be  
\label{eq_WIBQcc}
  \gt^{adme}_{\mu\alpha}(0,-k,0,k) =  g_{\mu\alpha} f^{amx}f^{dex}B_1(0,k,-k) + \cdots\,,
\ee 
where the ellipsis denotes terms proportional to  $k_\alpha$, 
which will be annihilated upon contraction with the 
projectors $P_{\nu}^{\alpha}(k)$ in  \1eq{d4_bbcc}. 
  
Substituting \1eq{eq_WIBQcc} into \1eq{d4_bbcc}, and  
employing the Jacobi identity together with the identity 
$f^{abe}f^{abx} = C_\mathrm{A} \delta^{ex}$,  
where  $C_\mathrm{A}$ is the Casimir eigenvalue of the adjoint representation [$N$ for SU($N$)], and using that $P_{\mu}^{\mu}(k)=3$, 
one arrives at the final result%
\begin{align}   
\label{eq_nova}
   \gh^{abmn}_{\mu\nu}(0,0,0,0) = g_{\mu\nu}(f^{max}f^{xbn} + f^{mbx}f^{xan})\left\{ Z_c + i\frac{3}{4}g^2C_{\rm A} \, Z_1 \! \int_k \! \Delta(k) D(k) B_1(0,k,-k) \right\} \,,
\end{align} 
or, equivalently, in terms of the form factor $T(0)$ defined in Eq.~\eqref{T0}
\begin{align}   
   T(0) = Z_c + i\frac{3}{4}g^2C_{\rm A} \, Z_1 \! \int_k \! \Delta(k) D(k) B_1(0,k,-k)  \,.
\label{eq_novat}
\end{align}

\begin{figure}[t]
    \centering
    \includegraphics[scale=0.7]{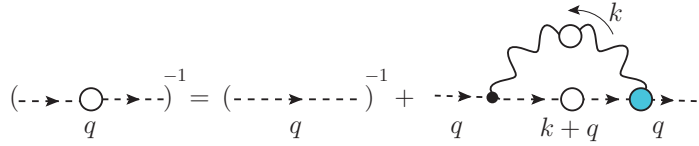}
    \caption{The SDE for the ghost propagator. The white circles represent the full gluon and ghost propagators, while the blue one denotes the full ghost-gluon vertex.}
    \label{fig_ghost}
\end{figure}

In order to make contact with the exact results of \2eqs{eq_bbccall}{T0}
we need to employ the  
SDE that governs the ghost dressing function $F(q)$ in the Landau gauge,
depicted in \fig{fig_ghost}. 
Using for $\gp_{\mu}(-q,k+q,-k)$ the decomposition given in    
\1eq{eq_B12}, the ghost SDE is given by    
%
%
\begin{align} 
     F^{-1}(q) = Z_{\rm c} + i g^2 C_{\rm A}Z_1 \int_k D(k+q) \Delta(k) f(k,q) B_{1}(-q,k+q,-k)\,,
\label{eq_ghr}
\end{align}
where the renormalization constants $Z_{\rm c}$ and $Z_1$ have been defined in \2eqs{renor_prop}{eq_z14}, respectively, and $f(k,q):= 1 - (k \vdot q)^2/k^2q^2$.

It is then elementary to demonstrate by setting $q=0$ into \1eq{eq_ghr}
that 
\be
F^{-1}(0) = T(0) \,.
\label{FT0}
\ee
Thus, as announced, the result of 
\eqref{eq_bbccall} is recovered from the SDE of the 
vertex $\gh^{abmn}_{\mu\nu}(q,r,p,t)$. 

We emphasize that throughout the entire demonstration leading to 
\1eq{eq_nova}, together with the subsequent 
equality in \1eq{FT0}, no approximations have been employed. 
The results obtained are therefore exact, and 
represent a rather special occurrence 
within the SDE formalism.

\section{Estimating truncation errors}    
\label{sec_numerics}

 In this section we explore 
 the implications of a simple truncation  
implemented at the level 
 of \1eq{eq_novat}, and  
estimate the associated errors by comparing the answers with the exact result 
of \1eq{T0}. The main points of this analysis may be summarized as follows. 

({\it i}) We approximate the ghost-gluon form factor $B_1(r,p,q)$ 
by its tree-level value, \ie we set $B_1(r,p,q)=1$, and determine the error induced 
by this simplification to $T(0)$, in two different situations: 

({\it a}) the ghost propagator $D(k)$  entering into 
\1eq{eq_novat} is {\it self-consistently} obtained from its own SDE, namely \1eq{eq_ghr} solved with $B_1(r,p,q)=1$, 

and  

({\it b}) the ghost propagator
$D(k)$ is treated as an {\it external input}, obtained from lattice 
simulations; it is  substituted into \1eq{eq_novat}, which is subsequently evaluated 
with $B_1(0,k,-k)=1$.

\begin{figure}[t]
    \includegraphics[width=0.45\textwidth]{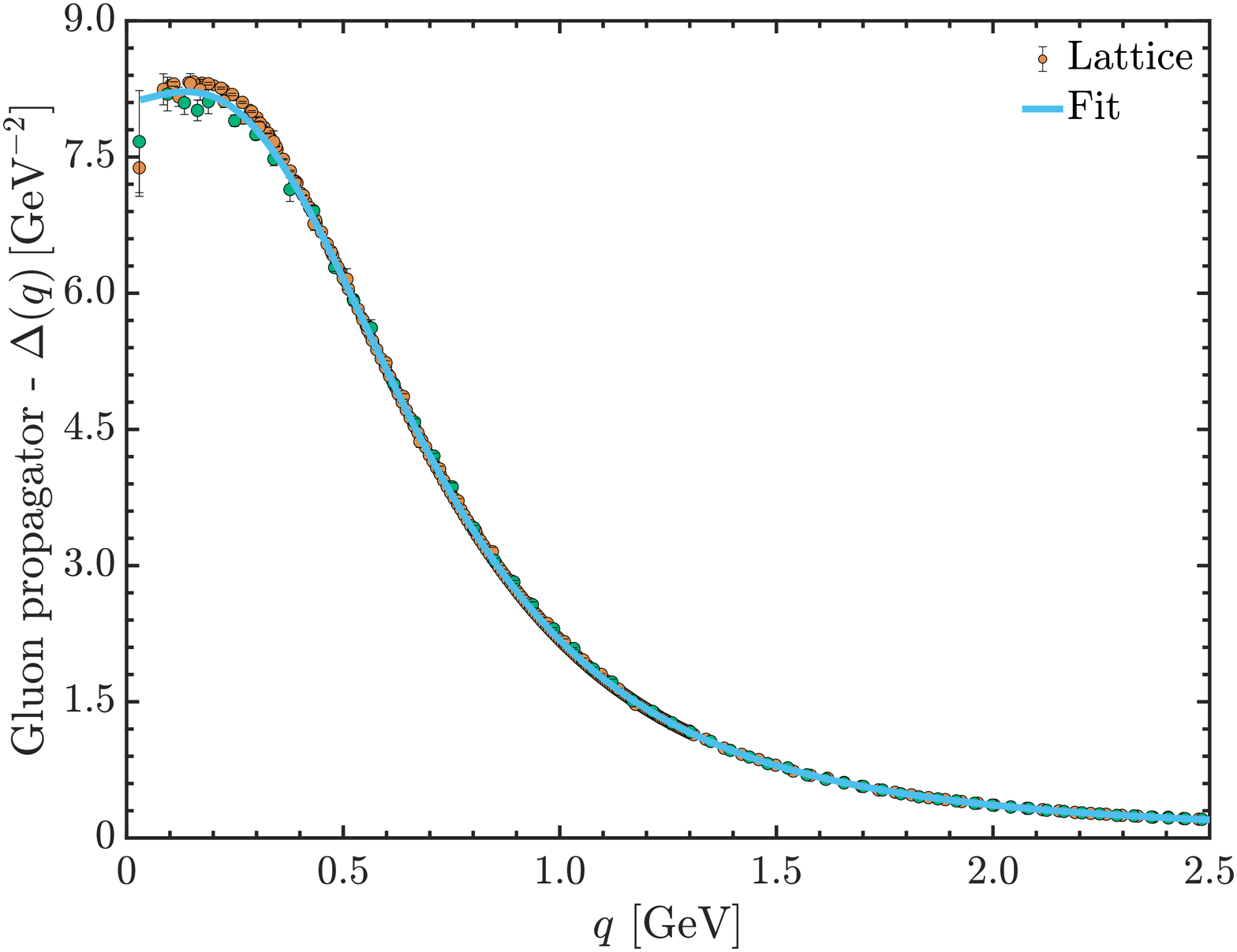} \quad
	\includegraphics[width=0.45\textwidth]{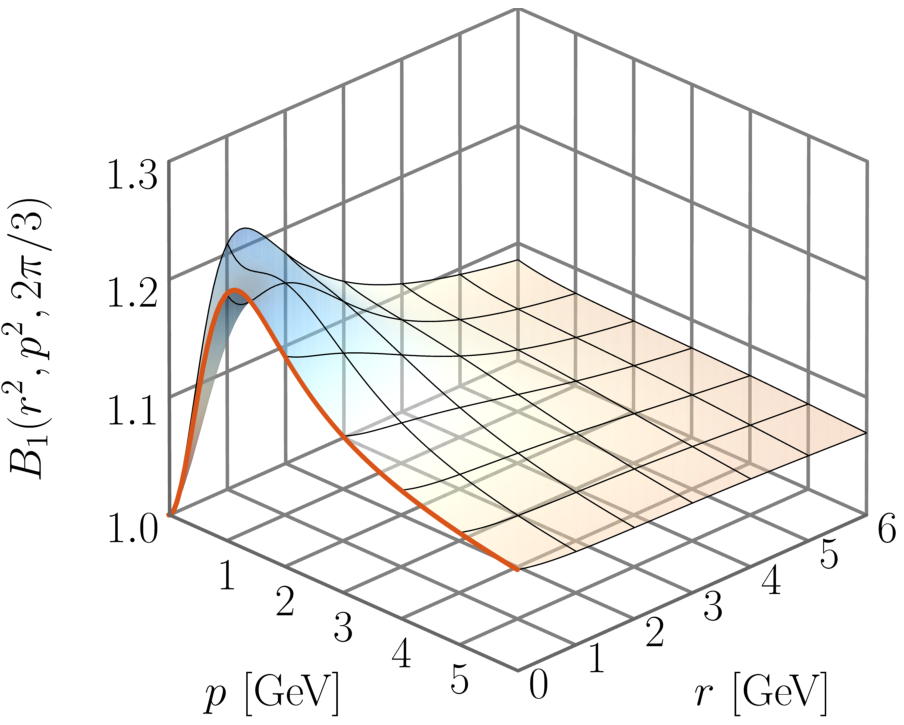}
 	\caption{{\it Left panel:} The lattice data of~\cite{Bogolubsky:2009dc,Boucaud:2017ksi,Boucaud:2018xup,Aguilar:2021okw} (circles), and the corresponding fit (blue continuous curve) for the gluon propagator, $\Delta(q)$. {\it  Right panel:} The ghost-gluon form factor $B_1(r^2,p^2,2\pi/3)$ in general kinematics for a fixed value of angle $\varphi=2\pi/3$. The orange curve highlights the soft antighost limit of $B_1(0,p,-p)$, entering in~\1eq{eq_novat}.}
\label{vertex}
\end{figure}

({\it ii}) Note that, in order to simplify the analysis, in both cases  
the gluon propagator $\Delta(k)$ will be an external input, 
obtained from the combined set of lattice data of~\cite{Bogolubsky:2009dc,Boucaud:2017ksi,Boucaud:2018xup,Aguilar:2021okw}; the corresponding 
fit, shown on the left panel of \fig{vertex}, is given by Eqs.~(B5) and~(B6) in~\cite{Aguilar:2021okw}.
We stress that the lattice data for both the gluon propagator and the ghost dressing function have been cured from discretization artifacts 
and finite-size effects~\cite{Duarte:2016iko,Boucaud:2017ksi,Boucaud:2018xup,Duarte:2017wte}.

({\it iii}) As far as the truncation errors are concerned,  
the main difference between the two cases is that in  
({\it a}) 
the error made when setting $B_1(r,p,q)=1$
affects the result for $T(0)$ nonlinearly, 
while in ({\it b})  the effect is practically linear.
Specifically, in  ({\it a}) the error induced 
to $T(0)$ by the corresponding error in 
$B_1$ is twofold: direct, through the slice $B_1(0,k,-k)$ 
in \1eq{eq_novat}, and indirect, 
through the entire $B_1(-q,k+q,-k)$ 
that enters in the SDE 
that determines $D(k)$ [see \1eq{eq_ghr}].
The difference between the two cases
is that in ({\it b})
the indirect error is eliminated, 
since $D(k)$ is fixed from the lattice, 
and coincides with the answer obtained from the full treatment of the ghost SDE.

({\it iv}) The result obtained from \1eq{eq_novat} for the cases ({\it a}) and ({\it b}),  
to be denoted by $T_{a}(0)$ and $T_{b}(0)$, respectively,
will be compared 
with the exact result $F^{-1}(0)$; its benchmark value,  
to be denoted by $F^{-1}_{\!\s L}(0)$, 
is taken from the lattice simulation of~\cite{Bogolubsky:2009dc,Boucaud:2018xup}.
The corresponding relative error, denoted by 
$\delta$, is defined as 
\begin{align}   
\label{eq_error}
\delta_{i} = \frac{\left| T_i(0) -F^{-1}_{\!\s L}(0)\right|}{F^{-1}_{\!\s L}(0)}\times 100 \% \,, \,\,\,\qquad i=a,b \,.
\end{align}
 %

({\it v}) All relevant quantities 
are renormalized using the momentum subtraction (MOM) 
scheme, where the renormalized two-point functions 
acquire their tree-level values at the subtraction  
point $\mu$, \ie  $\Delta^{-1}(\mu) = \mu^2$ and $F(\mu)=1$. 
Within MOM we employ the special case of the so-called Taylor scheme~\mbox{\cite{Boucaud:2008gn,Boucaud:2011eh,vonSmekal:2009ae}}, 
which imposes the additional condition that the renormalized ghost-gluon vertex 
reduces to tree-level in the soft ghost kinematics, \ie 
$\gp_\mu(r,0,-r) = r_\mu$. This last condition  
fixes the (finite) ghost-gluon renormalization constant at the special value \mbox{$Z_1=1$}. 
In our computations, 
the subtraction point is chosen to be  
\mbox{$\mu= 4.3$ GeV}; the corresponding value  
for \mbox{$\alpha_s(\mu) = g^2/4\pi$} 
is \mbox{$\alpha_s(4.3 \,\rm GeV) =0.244$}~\cite{Aguilar:2021okw}.

({\it vi}) 
It is instructive to compare the approximation 
\mbox{$B_1(r,p,q)=1$} with the full $B_1(r,p,q)$ in general kinematics, 
obtained from a detailed SDE study of the ghost-gluon vertex~\cite{Aguilar:2021okw}. To that end, 
in the right panel of \fig{vertex} we show a representative result for $B_1(r^2,p^2,\varphi)$, 
 where $\varphi$ is the angle between $r$ and $p$, and we choose 
 $\varphi=2\pi/3$. 
 When $r \approx p \approx \mbox{$1.2$~GeV}$,  
$B_1$ displays a moderate 
 peak of approximately $22\%$ over the tree-level value.
 On the other hand, 
 when both momenta vanish, $B_1$ 
 reduces to its tree-level value. 
 On the same plot we highlight by the continuous orange line the soft antighost limit $B_1(0,p,-p)$, entering in \1eq{eq_novat}. 
 
({\it vii}) For the implementation of the truncation  
mentioned in case ({\it a}), 
consider the ghost SDE 
in \1eq{eq_ghr} , with $Z_1=1$ in the Taylor scheme\footnote{Note that 
from $F(\mu)=1$ follows that 
$ Z_{\rm c} =1 - i g^2 C_{\rm A} \int_k D(k+\mu) \Delta(k) f(k,\mu) B_1(-\mu, k+ \mu, -k)$.}.
Clearly, for the numerical analysis,   
the Euclidean space versions  
of all expressions must 
be used; for the standard conversion rules, see \eg Eq.~(5.1) of~\cite{Aguilar:2018csq}. 

\begin{figure}
\begin{center}
 	\includegraphics[scale=0.25]{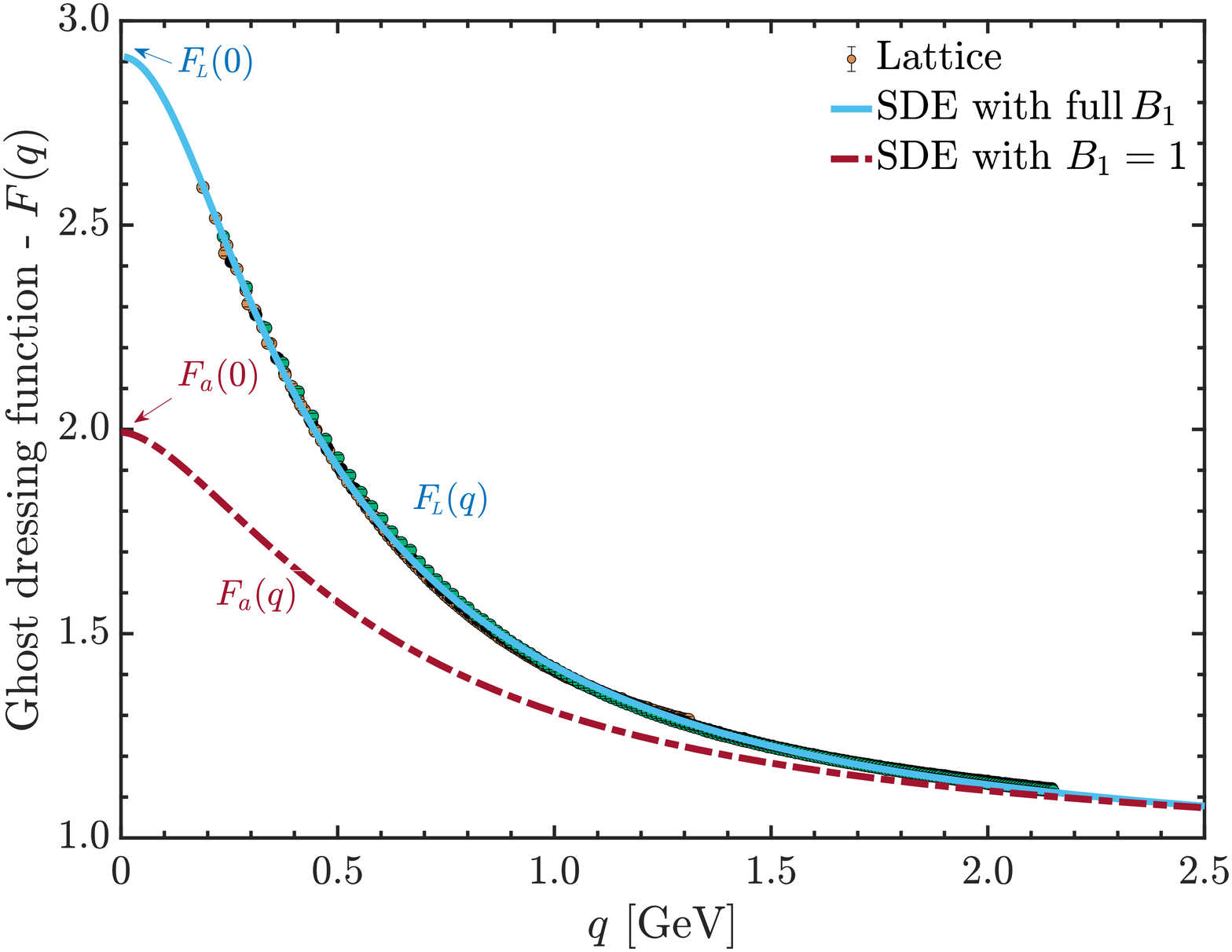} \quad
 	\includegraphics[scale=0.25]{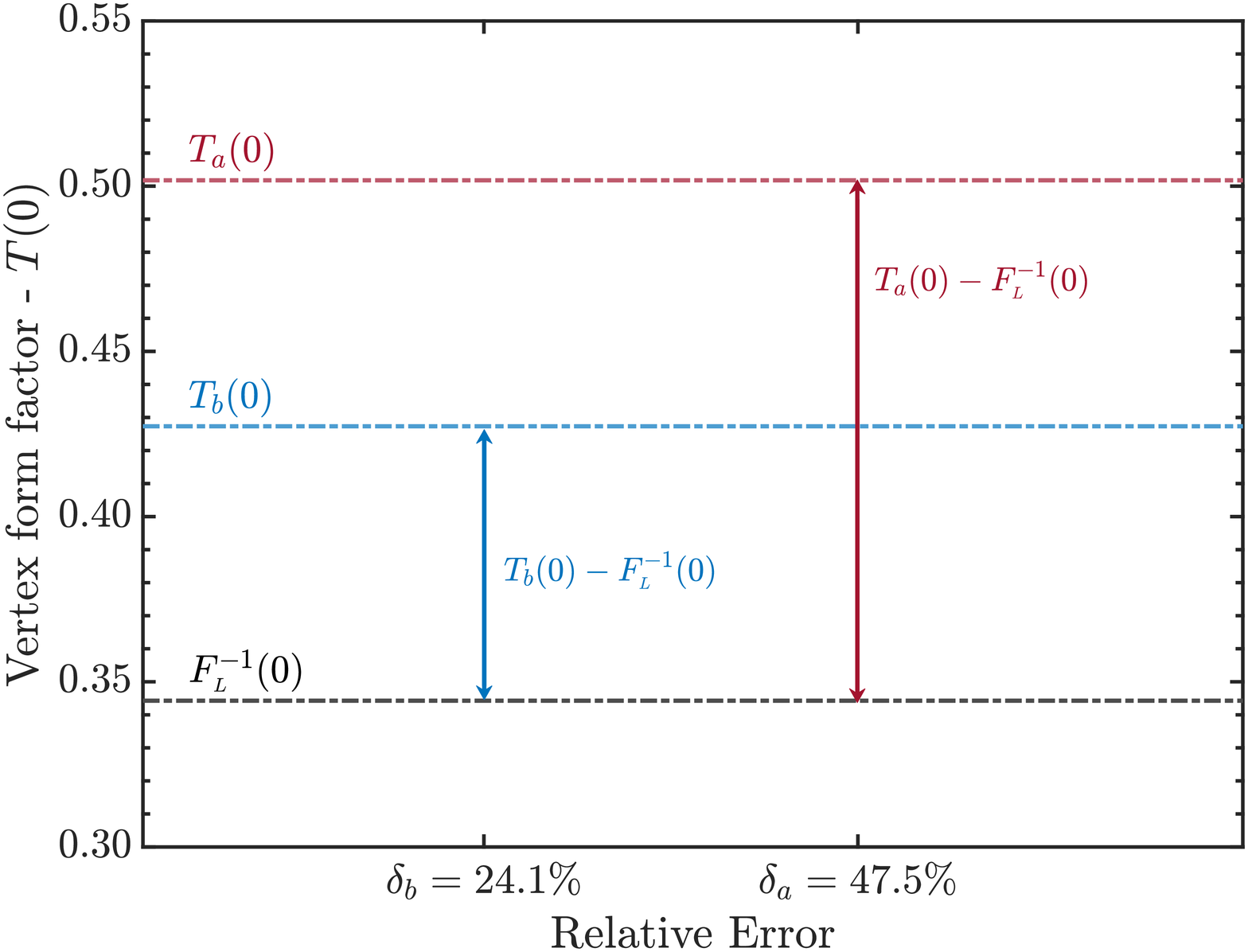}
 	\end{center}
 	\caption{{\it Left panel:} 
Lattice data for the ghost dressing function from \cite{Boucaud:2018xup} (circles), 
together with SDE results using the full $B_1$ (blue continuous curve) 
or its tree-level value (red dashed curve).
{\it Right panel:} Comparison of the exact value for  $T(0)$ with corresponding values obtained  when we use in the \1eq{eq_novat} the two curves for $F(q)$  shown in the left panel.}
\label{inputs}
\end{figure}

Let us first emphasize that 
when the complete $B_{1}(-q,k+q,-k)$ is employed as input, together with the 
$\Delta(k)$ mentioned in ({\it ii}) and the $\alpha_s(\mu)$ in ({\it v}), 
\1eq{eq_ghr} returns a solution for $F(q)$ 
that is in excellent agreement with the 
lattice data of~\cite{Boucaud:2018xup}, see blue continuous curve in the left panel of \fig{inputs}~\cite{Aguilar:2021okw}. 
From this curve we may directly deduce that the 
benchmark value is given by $F^{-1}_{\!\s L}(0) = 0.344$ (for \mbox{$\mu= 4.3$ GeV}). 

Next, we implement the approximation $B_{1}(-q,k+q,-k)=1$ 
in \1eq{eq_ghr}, 
keeping $\Delta(k)$ and $\alpha_s(\mu)$ fixed. 
The resulting integral equation is given by 
\begin{align} 
F^{-1}(q) = i g^2 C_{\rm A} \int_k \Delta(k) \left[D(k+q) f(k,q) - D(k+\mu)f(k,\mu)\right]\,,
\label{gap1}
\end{align}
and the corresponding solution, to be denoted by $F_a(q)$,   
is shown as the red dashed curve in  the left panel of \fig{inputs}.
It is clear that $F_a(q)$ deviates considerably from the 
lattice results; in particular, at the origin we have the value 
$F^{-1}_{a}(0) = 0.501$. Then, by virtue of \1eq{FT0}, we have that 
$T_{a}(0) = F^{-1}_{a}(0)$, 
and employing \1eq{eq_error} we find that the corresponding relative error is given by  
$\delta_{a} = 47.5\%$.


({\it viii}) 
We now turn to the truncation of case ({\it b}). 
In order to obtain the relative error we simply need to evaluate  
the integral of \1eq{eq_novat} with $B_1(0,k,-k)=1$, but 
setting $D(q)\to D_{\!\s L}(q) = F_{\!\s L}(q)/q^2$, namely the lattice (and full SDE) result 
for the ghost propagator. Thus, 
the integral to be determined is given by 
\begin{align} 
T_b(0) = 
i g^2 C_{\rm A} \int_k \Delta(k) \left[D_{\!\s L}(k+q) f(k,q) - D_{\!\s L}(k+\mu)f(k,\mu)\right]\,.
\label{nogap}
\end{align}
It is clear that this ``hybrid'' treatment invalidates the equality of 
\1eq{FT0}, since the r.h.s. of \1eq{nogap} does {\it not} coincide with the $q=0$ limit 
of any ghost SDE; thus, $T_b(0) \neq F_{\!\s L}(0)$.   
In particular, 
using in \1eq{nogap} the same $\Delta(k)$ and $\alpha_s(\mu)$ as before, together 
with a fit to the lattice data for $F_{\!\s L}(q)$, 
one obtains $T_b(0) = 0.427$, and the associated relative error computed from  
\1eq{eq_error} is $\delta_{b} = 24.1\%$.

 ({\it ix})  The results derived in this section are pictorially summarized in the 
 right panel of \fig{inputs}. It is clear that case ({\it b}) 
 leads to a considerably smaller error, since the treatment of the 
 ghost dressing function as an external input linearizes 
 the dependence of $T(0)$ on $B_1$. In fact, the relative error $\delta_{b} = 24.1\%$ 
 is very close to the $22.1\%$ that separates the peak in $B_1$ from the tree-level value [see \fig{vertex} and comments in ({\it vi})]. 
 This notable reduction in the error 
 with respect to case ({\it a}) 
 suggests that the hybrid treatment may be preferable, albeit theoretically less rigorous.

\section{Discussion and Conclusions}
\label{sec:conc}

In the present study we have considered 
the SDE of the special four-particle vertex
that controls the interaction of two background 
gluons with a ghost-antighost pair 
(\bbcc vertex). 
We focused on the deep 
infrared limit,  
where all incoming momenta vanish.
In this extreme limit, the form factor of the 
vertex may be determined exactly
by virtue of the Abelian STI that it satisfies.
In particular,  
the two surviving form factors are expressed  
in terms of two- and three-point functions,
but without contributions from ghost-gluon kernels. 
This central result is recovered from the SDE 
by exploiting the Abelian STIs of the 
various vertices nested inside the diagrammatic 
expansion, and making repeated use of 
Taylor's theorem. 
The aforementioned exact result, in turn, allows for 
the determination of the error associated 
with two related, but inherently distinct 
truncation procedures. 

The realization of the WI  
at the level of the SDE, presented in Sec.~\ref{sec_WTIbbcc}, is particularly noteworthy, 
combining concepts and techniques in a novel way,  
not previously presented in the literature. 
This construction highlights the importance of incorporating 
into the SDEs vertices  
that satisfy the constraints 
imposed by the fundamental symmetries of the theory, such as the STIs and Taylor's theorem. 
In addition, it exposes the tight connections and delicate balance among the various 
SDE components, necessary for maintaining certain basic relations intact.

In this context, note that in \1eq{eq_novat}, 
the dependence of 
$T(0)$ on the form factor $B_1$ of the ghost-gluon vertex $\gp_{\mu}(r,p,q)$ 
does not originate from the graphs ($a_1$), 
($a_2$), ($a_3$), ($a_4$), ($a_5$), ($b_1$), and ($b_2$) in \fig{fig_completed},  
which contain the $\gp_{\mu}(r,p,q)$ explicitly, 
since they all vanish in the all-soft limit, see Sec.~\ref{sec:SDE_deriv}. 
Instead, the answers stem from graphs ($a_7$) and ($b_3$),  
which have no explicit dependence on this form factor, but contain the vertex 
$\gt_{\mu\alpha}^{adme}(0,-k,0,k)$, whose WI 
in  \1eq{eq_WIBQcc} induces the final dependence of $T(0)$ on $B_1$.

SDE calculations are often simplified 
by using lattice results as inputs for certain basic Green's functions,
and one of the truncations considered in this study 
[case ({\it b}) in item ({\it i}) of Sec.~\ref{sec_numerics}]
is motivated by this particular practice. 
In the present analysis, we found that this procedure clearly lowers  
the error induced to $T(0)$ by the uncertainty in $B_1$. 
To be sure, the actual amount of error reduction achieved 
may be atypically high (a factor of two), owing mostly 
to the enhanced sensitivity of the ghost SDE on $B_1$.
An analogous study at the level of the 
gluon propagator is likely to produce a  
milder dependence on $B_1$, and thus, a more moderate improvement; on the other hand, 
the SDE of the gluon propagator 
depends on other poorly known Green's functions, such as the four-gluon vertex~\cite{Driesen:1998xc,Kellermann:2008iw,Binosi:2014kka,Cyrol:2014kca,Eichmann:2015nra,Huber:2017txg,Huber:2020keu,Catumba:2021qbh}. Thus, 
even though no rigorous conclusions may be drawn, 
the use of lattice inputs in SDEs emerges as an advantageous option, because
it reduces the number of 
active coupled equations and lowers the overall error. 
In addition, 
advances in techniques and procedures
used for the elimination of discretization artifacts and finite-size effects from the data~\cite{Duarte:2016iko,Boucaud:2017ksi,Boucaud:2018xup,Duarte:2017wte}
put the synergy between (gauge-fixed) 
lattice simulations and SDEs on firmer theoretical ground.

Undoubtedly, the BFM provides an excellent testing ground for 
the set of ideas presented in this work, mainly due to the ghost-free 
STIs satisfied by the relevant Green's functions. 
It would be interesting to carry out 
lattice simulations directly in the BFM~\cite{Dashen:1980vm,Luscher:1995vs}, following the 
formalism introduced in~\cite{Cucchieri:2012ii,Binosi:2012st}.

\section{Acknowledgments}
\label{sec:acknowledgments}
The work of  A.~C.~A. and B.~M.~O. are supported by the CNPq grants 307854/2019-1 and 141409/2021-5, respectively.
A.~C.~A also acknowledges financial support from the FAPESP project 2017/05685-2 and 464898/2014-5 (INCT-FNA).
J.~P. is supported by the Spanish MICINN grant PID2020-113334GB-I00 
and the regional Prometeo/2019/087 from the Generalitat Valenciana.

\newpage
\appendix
\section{Feynman rules for BFM vertices}
\label{sec:App_feynman}

\begin{table}[h]
\setlength{\extrarowheight}{5pt}
\begin{tabularx}{\linewidth}{|c|X|}
\hline
\includegraphics[scale=0.45]{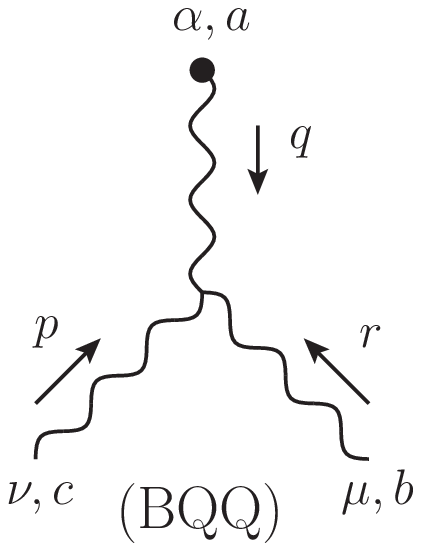} & \vspace{-3.0cm} 
\begin{equation}
\label{BQQ0}
    \gt_{\alpha\mu\nu}^{(0)}(q,r,p) = g_{\mu \nu}(r-p)_{\alpha} + g_{\alpha \nu}(p-q)_{\mu} +  g_{\alpha \mu}(q-r)_{\nu} + \xi^{-1} (g_{\alpha\nu} r_\mu - g_{\alpha\mu} p_\nu) \,,  \vspace{-5cm}
\end{equation}  \\ \hline
\includegraphics[scale=0.5]{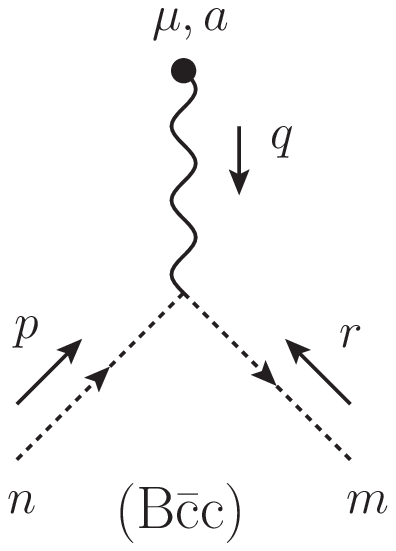} & \vspace{-3.0cm}
\begin{equation}
\label{BCC0}
    \gt^{(0)}_\mu(r,p,q)=(r-p)_\mu \,, \vspace{-4cm}
\end{equation} \\
\hline
\includegraphics[scale=0.5]{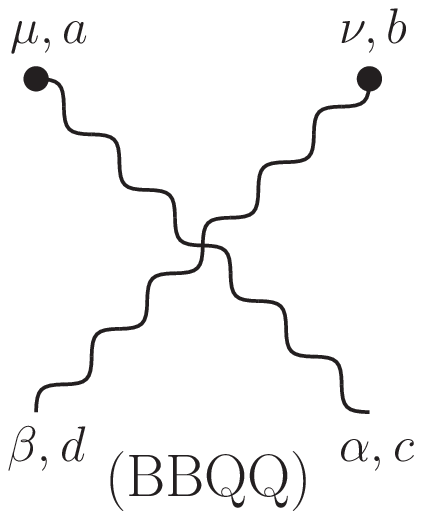} & \vspace{-3.4cm}
\begin{eqnarray}
	\label{BBQQ0}
    &&\gh^{(0)abcd}_{\alpha\beta\mu\nu}= f^{adx}f^{xcb}\left(g_{\alpha\mu}g_{\beta\nu}-g_{\alpha\beta}g_{\mu\nu}\right) +f^{abx}f^{xdc}\left(g_{\alpha\nu}g_{\beta\mu}-g_{\alpha\mu}g_{\beta\nu}\right)  \\
    &+&f^{acx}f^{xdb}\left(g_{\alpha\nu}g_{\beta\mu}-g_{\alpha\beta}g_{\mu\nu}\right) + \xi^{-1}(f^{adx}f^{xbc}g_{\alpha\nu}g_{\beta\mu}-f^{acx}f^{xdb}g_{\alpha\mu}g_{\beta\nu}),\nonumber  \vspace{-7cm}
\end{eqnarray} \\
\hline
\includegraphics[scale=0.5]{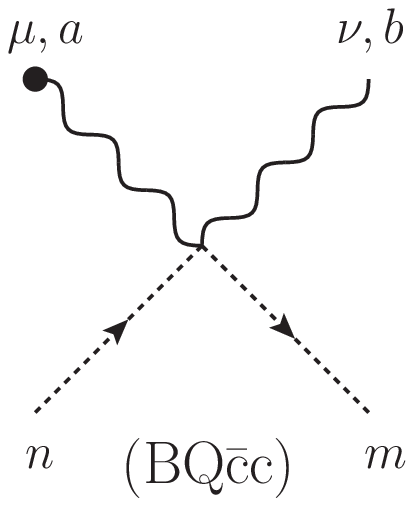} & \vspace{-2.7cm}
\begin{equation}
 \label{BQcc0}
    \gt^{(0)abmn}_{\mu\nu}=f^{max}f^{xbn}g_{\mu\nu} \,, \vspace{-5.5cm}
\end{equation} \\
\hline
\includegraphics[scale=0.5]{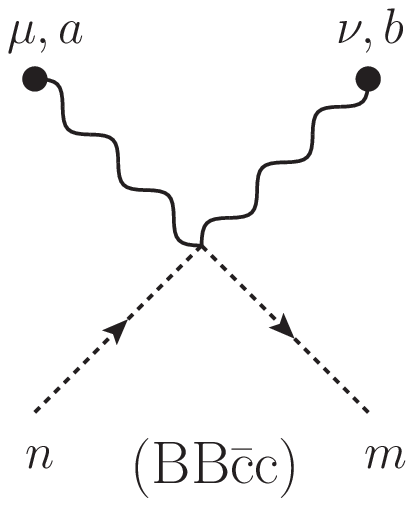} & \vspace{-2.7cm}
\begin{equation}
\label{BBcc0}
    \gh^{(0)abmn}_{\mu\nu}=g_{\mu\nu}\left(f^{max}f^{bnx}+f^{mbx}f^{anx}\right) \,. \vspace{-5.5cm}
\end{equation} \\
\hline
\end{tabularx}
\caption{The diagrammatic representations of the new vertices appearing in the BFM and their respective Feynman rules at tree-level~\cite{Binosi:2009qm}. Notice that for the three-point functions we have factored out the coupling $g$ and their respective color structure,  following the definitions of Eq.~\eqref{def3g}, while for the four-point functions,  we have factored out only $-ig^2$ as shown in Eq.~\eqref{def4g}.}
\label{fig_feyback}
\end{table}

\newpage
%

\end{document}